\documentclass[conference,onecolumn]{IEEEtran}
\IEEEoverridecommandlockouts

\usepackage{cite}
\usepackage{amsmath,amssymb,amsfonts}
\usepackage{algorithm}%
\usepackage{algorithmicx}%
\usepackage{algpseudocode}%
\usepackage{graphicx}
\usepackage{textcomp}
\usepackage{xcolor}
\usepackage{tabularx, booktabs}
\usepackage{caption}
\usepackage{mathtools}
\usepackage{orcidlink}

\usepackage{booktabs}   
\usepackage{graphicx}   
\usepackage{pdflscape} 
\usepackage{rotating}
\usepackage{adjustbox}
\usepackage{longtable}

\def\BibTeX{{\rm B\kern-.05em{\sc i\kern-.025em b}\kern-.08em
    T\kern-.1667em\lower.7ex\hbox{E}\kern-.125emX}}
\begin{document}

\title{Divide-et-impera Heuristic-based Randomized Search for the Qubit Routing
Problem
}


\author{%
 \IEEEauthorblockN{Marco Baioletti \orcidlink{0000-0001-5630-7173}\textsuperscript{*}}
 \IEEEauthorblockA{Department of Mathematics and Computer Science\\
 University of Perugia\\
 Perugia, Italy\\
 marco.baioletti@unipg.it}
 \and
 \IEEEauthorblockN{Fabrizio Fagiolo \orcidlink{0009-0003-0390-7855}\textsuperscript{*}}
 \IEEEauthorblockA{Institute of Cognitive Sciences and Technologies (ISTC)\\
 National Research Council (CNR)\\
 Rome, Italy\\
 fabrizio.fagiolo@istc.cnr.it}
 \and
 \IEEEauthorblockN{Angelo Oddi \orcidlink{0000-0003-4370-7156} \textsuperscript{*}}
 \IEEEauthorblockA{Institute of Cognitive Sciences and Technologies (ISTC)\\
 National Research Council (CNR)\\
 Rome, Italy\\
 angelo.oddi@istc.cnr.it}
 \and
 \IEEEauthorblockN{Riccardo Rasconi \orcidlink{0000-0003-2420-4713}\textsuperscript{*}}
 \IEEEauthorblockA{Institute of Cognitive Sciences and Technologies (ISTC)\\
 National Research Council (CNR)\\
 Rome, Italy\\
 riccardo.rasconi@istc.cnr.it}
 \thanks{\textsuperscript{*}\;All authors contributed equally to this research.}
 }

\maketitle

\begin{abstract}



This paper introduces the DIRSH algorithm for the Qubit Routing Problem (QRP), using a heuristic-guided randomized divide-and-conquer strategy. The method splits the circuit into chunks and optimizes each one with a stochastic selection of gates and swaps. It balances global search, via restarts and adaptive tuning of bandit parameters with depth-sensitive local pruning.

Tested on RevLib benchmarks mapped to the 20-qubit IBMQ Tokyo topology, DIRSH outperformed three LightSABRE variants across different time budgets, achieving shorter depths and fewer swaps. These results confirm that combining chunk-based decomposition with bandit-driven heuristics is effective for routing quantum circuits on NISQ devices.
\end{abstract}


\section{Introduction}
    \label{sec:introduction}

Quantum algorithms are expressed using quantum circuits, which are characterized by sequences of quantum gates applied to physical qubits according to the architecture of a quantum processor.
However, executing these circuits on real NISQ (Noisy Intermediate-Scale Quantum) devices \cite{preskill2018quantum} presents significant challenges due to the presence of both hardware constraints and decoherence effects.

One of the main challenges in current quantum technologies is the limited qubit connectivity, where only certain pairs of qubits can interact directly, thus limiting the applicability of binary quantum gates to neighboring qubits only.
As a result, when a two-qubit gate involves qubits that are not physically adjacent in the quantum physical back-end, additional routing strategies are needed to enable their interaction. 
This problem is known as the Qubit Routing Problem (QRP) \cite{theory-of-QRP}, whose objective is to find a mapping between logical and physical qubits such that each two-qubit gate in the circuit operates on physically connected qubits at all times.

Quantum compilers achieve this by inserting appropriate SWAP gates, which exchange the quantum states (a.k.a. logical qubits) of adjacent qubits (i.e., the physical qubits). 
In this way, the logical qubits that need to interact are brought closer together in the physical layout whenever necessary for the circuit's execution.

The primary goal of quantum compilers is to minimize the number of \textit{swap gates}, since each one introduces significant additional noise, thus increasing the probability of errors.
At the same time, it is crucial to limit the circuit \textit{depth}, that is, the number of layers of sequential gates, since deeper circuits are more susceptible to decoherence.

Currently, the most effective methods in the literature to solve the qubit routing problem are SABRE \cite{SABRE} and its improved version \textit{LightSABRE} \cite{lightSabre}. 
The algorithm we present in this work, Divide-et-Impera Randomized Search with Heuristics (\textit{DIRSH}) is based both on techniques already established in the literature and on innovative approaches more recently developed to improve the results for the QRP. 
\textit{DIRSH} is designed to address the qubit routing problem on general quantum circuits, composed of both unary (one-qubit) and binary (two-qubit) gates. 

The rest of the paper is organized as follows.  
Section~\ref{sec:related_work} describes the related work, while Section~\ref{sec:problem_definition} proposes a formal definition of the solved problem. Section~\ref{sec:algorithm} introduces the proposed solving algorithm. 
Subsequently, Section~\ref{sec:evaluation} describes and comments the obtained experimental results, followed by Section~\ref{sec:conclusions} which ends the paper by providing some concluding remarks and some possible lines of future research work.
Section~\ref{sec:additional-material} reports the results of the algorithm in more detail.














\section{Related Work}
     \label{sec:related_work}

The Qubit Routing Problem (QRP) involves transforming a logical quantum circuit into a sequence of operations that are compatible with the physical qubit topology, while minimizing either the overall circuit depth (makespan) or the number of additional SWAP gates required.
QRP has been addressed in the literature using a variety of approaches, ranging from symbolic planning to metaheuristic techniques.

In~\cite{Venturelli_2018} the compilation problem is formulated as a temporal planning problem, where each gate is modeled as a PDDL action with preconditions and effects defined on the qubits, and general-purpose temporal planners are used to produce minimum-duration plans. This method has been shown to generate optimal solutions, although it is limited to small circuits due to the combinatorial explosion of the search space.

To improve scalability, in~\cite{greedy_randomized} a greedy randomized heuristic is proposed, based on a lexicographic ranking function. A global indicator (closure) guides the reduction of the residual makespan, while a local tie-breaker prevents relocation cycles. Thanks to its randomized component, the algorithm is able to quickly produce high-quality solutions and has proven competitive on the tested benchmarks.

The exploration is further advanced in~\cite{aco_compilation}, where the authors introduce \textit{QCC-ACO}, an Ant Colony Optimization variant. In this approach, a pheromone model is constructed directly on the dependency graph of the gates, while the priority rule for reasoning on each insertion step leverages a greedy ranking strategy. \textit{QCC-ACO} achieves several new best makespan results across a large set of benchmark instances.

A further advancement is presented in \cite{chand}, where two rollout-based heuristics are proposed. The first employs sequential decision-making guided by a priority rule that estimates the makespan, while the second alternates between the rollout heuristic and a simple heuristic rule to broaden the exploration of the search space.

Another notable contribution is \cite{ARUFE1}, where genetic algorithms are used to produce the current best results against the used benchmark. 
In \cite{ARUFE1} the authors introduce the Decomposition-Based Genetic Algorithm (\textit{DBGA}), which breaks down QAOA-MaxCut circuits into sequential rounds.
Each round is addressed by evolving a population of partial solutions, which are passed to the next round through a decoder that inserts only the essential swap operations. 
This approach significantly reduces the search space, matching and at times even improving the current state-of-the-art performance.

In a subsequent study \cite{Arufe2}, the same authors introduce a novel genetic encoding/decoding scheme that directly represents the logical positions of gates. During decoding, only the minimum necessary swap operations are inserted, extending the algorithm's applicability to QAOA circuits for MaxCut. 
This approach achieves a reduction in average circuit depth compared to \textit{DBGA}, while maintaining similar convergence times.

Although the contributions discussed so far primarily target QAOA-related circuits, where quantum gates are typically commutative and can be reordered freely during routing, the underlying genetic encoding schemes and heuristic functions rely on the same unary and binary gates that are present in general-purpose quantum circuits. This modular design makes them easily adaptable to the compilation of general-purpose circuits, where the only difference is that gate dependencies follow a fixed partial order that must be preserved during compilation.

Besides the compiling procedures specialized for QAOA-related circuits discussed above, a number of compilers for general-purpose circuits have also been developed to integrate routing, depth optimization, and swap minimization strategies. 
Notable examples include t$|$ket$>$\cite{Tiket}, \textit{SABRE }\cite{SABRE}, and its lightweight variant \textit{LightSABRE} \cite{lightSabre}, whose empirical results show substantial improvements in the minimization of both the circuit depth and the number of the additional swap gates, outperforming several competing approaches.
An additional line of research is presented in \cite{Nannicini}, that explicitly formulates the qubit routing problem as an integer linear program where binary variables encode both the initial qubit allocation and the routing operations. The problem is solved with exact optimization methods, showing that linear programming techniques can effectively capture both the mapping and routing constraints.















\section{Problem definition}
    \label{sec:problem_definition}
An instance of the \textit{Quantum Routing Problem} (QRP) is defined as a tuple $P=\langle C_{0}, QM\rangle$, where $C_{0}$ denotes the logical quantum circuit to be compiled and $QM$ is a graph-based abstraction of the target quantum processor.

  $C_0$ is defined by a tuple $\langle  LQ, V_0, TC_0 \rangle$, where: 
  \begin{itemize}
    \item $LQ=\{q_{1},q_{2},\dots ,q_{N}\}$ is the set of \textit{logical qubits}.

    \item $V_{0}$ is the set of elementary quantum gates (unary or binary) that actually appear in the original circuit, without assuming a specific set of universal gate.
    Each gate $g\in V_{0}$ is uniquely identified by its type and the logical qubits on which it acts.
    
    \item $TC_{0}$ is the set of precedence constraints that preserve the execution order specified by the original circuit description.

\end{itemize}

The quantum machine is modeled as an undirected graph $QM=\langle PQ,E_{bin}\rangle$, where the vertices $PQ$ represent the physical qubits and the edges $E_{bin}$ specify pairs of qubits that can interact directly.
From $E_{bin}$ it is possible to define the set $\mathcal{SW}$ of all possible swap gates
$swap(q_i,q_j)$, such that $(q_i,q_j)\in E_{bin}$.


A solution is described by the tuple $S=\langle \mathit{SWAP}, TC\rangle$, where
\begin{enumerate}
  \item $\mathit{SWAP}$ is the set of additional swap gates required to satisfy the adjacency condition for all binary gates;
  \item $TC$ is a set of supplementary temporal precedence constraints such that:
        \begin{itemize}
          \item for every logical qubit $q_i$ a total order $\preceq_i$ is imposed on the set $Q_i$ of gates acting on $q_i$;
          \item every binary gate is mapped to a pair of physical qubits that are adjacent in $QM$;
          \item the dependency graph $\langle V_S,TC_S\rangle$ with $V_S = V_{0}\cup\mathit{SWAP}$ and $TC_S = TC_{0}\cup TC$ is acyclic.
        \end{itemize}
\end{enumerate}

Each gate $g\in V_S$ is assigned a level $\ell(g)$ computed with the following formula

$$ \ell(g)=1+\max_{g' \text{ precedes } g} \ell(g')$$

If there is no gate preceding $g$, then $\ell(g)=0$.

Given the previous definitions, two different objective functions can be used to define an optimal solution for the QRP. 

The \textit{depth} $de(S)$ of a feasible solution $S$ is the maximum of $\ell(g)$ for each gate $g$ appearing in $S$.
A depth‑optimal solution $S^*_{de}$ is a feasible solution that achieves the minimum depth among all feasible solutions.

The second objective function is the number of swap gates $sw(S)$ present in $S$.
A swap‑optimal solution $S^*_{sw}$ is a feasible solution that requires the minimum number of swaps among all feasible solutions.

\section{Proposed Algorithm}
    \label{sec:algorithm}

In this section we describe the proposed algorithm which
solves the QRP $\langle  C_0,  QM \rangle$ by minimizing the objective function $f_{obj}$, which can be either the number of swap gates $sw$ or the depth $de$.
The algorithm is similar for both objective functions; there are only a few differences between the two modes.

We need the concept of qubit assignment, which is a mapping $\sigma:LQ\to PQ$ that indicates for each logical
qubit the corresponding physical qubit. 
The assignment $\sigma$ is updated every time a new swap gate is added to the solution.

Basically, the algorithm has a time budget of $T$ seconds and produces the best solution found according to $f_{obj}$.

The circuit can be divided into $n_c$ chunks of the same depth (except the last one, which contains all remaining gates), and the algorithm calls the \textsc{Optimize\_Chunk} function on each chunk for $\Delta=\lfloor T/n_c \rfloor$ seconds, as shown in Algorithm \ref{alg:optimize}. 
The initial assignment $\sigma$ is set to a default allocation, where each
logical qubit is assigned to the homonymous physical qubit.
The solutions of each chunk are concatenated and the final assignment is used as the initial assignment for the next chunk.

\begin{algorithm}[t]
	\centering
	\caption{\textsc{Optimize}\label{alg:optimize}}
	\begin{algorithmic}
	    \Require {A problem $P = \langle  C_0,  QM \rangle$}
	    \State $\sigma \gets$ default initial assignment
        \State $\Delta t \gets \lfloor T/n_c \rfloor$
        \State $s \gets$ empty solution
	    \For{$j \gets 1$ to $n_c$}
        \State $C_j \gets j$-th chunk of $C_0$
        \State $s_j \gets$\textsc{Optimize\_Chunk}$(C_j, \sigma, \Delta t)$
        \State $s\gets$ concat$(s,s_j)$
        \State $\sigma\gets$ final state of $s_j$
	    \EndFor
		\State \Return $s$
	\end{algorithmic}
\end{algorithm}

The compilation of each chunk runs (function \textsc{Optimize\_Chunk})
for at most $\Delta t$ seconds, repeatedly calling the function \textsc{Generate\_Solution} and updating the best solution $s_{best}$
i.e., the incumbent solution found so far (not necessarily the global optimum).

The generation of a solution for a chunk is described in Algorithm \ref{alg:generate}.
The function initializes $s$ with a partial copy of $s_{best}$, i.e. 
$$s \gets \{\, g \in s_{best} : \ell(g) \le \frac{1}{2}de(s_{best}) \,\}$$ 
with a low probability of $10\%$;
otherwise, $s \gets \emptyset$.

Any binary gate $g$ in $C_j$ which is not yet in $s$ can be in three different states:
(i) $g$ is \textit{not-supported} if some of the gates preceding $g$ in $C_j$ are not present in $s$; (ii) $g$ is \textit{executable} if the logical qubits involved in $g$
lie on connected physical qubits according to the current qubit assignment $\sigma$; (iii) otherwise $g$ is only \textit{supported}.
If $g$ is a unary gate, it is always \textit{executable}.
Similarly, all possible swap gates in $\mathcal{SW}$ are always \textit{executable}.
A solution $s$ is complete when all the gates in $C_j$ appear in $s$.

The set of candidate gates $\mathcal{C}$, as required in the function \textsc{Generate\_Solution}, is computed in the following way.
If $f_{obj}$ is $sw$, $\mathcal{C}$ contains all the gates whose state is \textit{executable}, including all the swap gates in $\mathcal{SW}$.
\begin{algorithm}[t]
	\centering
	\caption{\textsc{Generate\_Solution} \label{alg:generate}}
	\begin{algorithmic}
	    \Require {A subproblem $P = \langle  C_j,  QM \rangle$, initial assignment $\sigma$, best solution $s_{best}$}
	    \If{random$<0.1$}
            \State $s\gets$ partial copy of $s_{best}$
        \Else
            \State $s\gets$ empty solution
        \EndIf
	    \While{$s$ is not completed}
        \State Find $\mathcal{C}$ the set of all the candidate gates
        \State Evaluate the heuristic functions $d_{sum}$ and $d_{min}$ on $\mathcal{C}$
        \State Prune $\mathcal{C}$
        \State Find $(\beta_1,\beta_2)$ using the multi-armed bandit method
        \State $g\gets $ Roulette\_Wheel$(\mathcal{C}, \beta_1, \beta_2)$
        \State Add $g$ to $s$
	    \EndWhile
		\State \Return $s$
	\end{algorithmic}
\end{algorithm}
If $f_{obj}$ is $de$, an index $t$ is used, which denotes the level of the
last gate added to $s$, and $\mathcal{C}$ is restricted to all executable gates $g$ which operate on physical qubits not used by gates at level $t$.
If there are no gates which satisfy this condition, $t$ is increased by $1$
and $\mathcal{C}$ is computed for the new value of $t$.

Two heuristic functions $d_{sum}$ and $d_{min}$ (first introduced in \cite{greedy_randomized}) are used to evaluate the gates in $\mathcal{C}$.
Such heuristic functions depend in their turn on the two functions $D_{sum}$ and $D_{min}$ for a qubit assignment $\sigma$, defined as
$$ D_{sum}(\sigma)=\sum_{(q_i,q_j)\in Q} d(\sigma(q_i),\sigma(q_j))$$
and
$$ D_{min}(\sigma)=\min_{(q_i,q_j)\in Q} d(\sigma(q_i),\sigma(q_j))$$

where $Q$ is the set of all pairs $(q_i,q_j)\in LQ\times LQ$, such that there exists a binary gate $g(q_i,q_j)$, which is supported and acts on qubits $q_i$ and $q_j$.

The heuristic functions of a swap gate $g$ which produces a new qubit assignment $\sigma'$   are defined as
$d_{sum}(g)=D_{sum}(\sigma')$ and $d_{min}(g)=D_{min}(\sigma')$.
For a non swap gate $g$, these functions are defined as
$d_{sum}(g)=D_{sum}(\sigma)$ (if $g$ is unary) or
$d_{sum}(g)=D_{sum}(\sigma)-1$ (if $g$ is binary), and $d_{min}(g)=1$, as described in \cite{greedy_randomized}.

The pruning method suggested in \cite{chand} is applied to the swap gates in $\mathcal{C}$, by keeping only the swap gates $g$ such that either $d_{sum}(g)<D_{sum}(\sigma)$ or 
$d_{sum}(g)=D_{sum}(\sigma) \wedge d_{min}(g)=D_{min}(\sigma)$.
Lastly, a probability value $p(g)$ is associated to each gate $g\in \mathcal{C}$ and defined as
$$ p(g) \sim (1-\bar d_{sum}(g))^{\beta_1} \cdot (1-\bar d_{min}(g))^{\beta_2}, $$
where $\bar d_{sum}(g)$ and $\bar d_{min}(g)$ are the values of the same functions normalized in $[0,1]$, while $\beta_1$ and $\beta_2$ are two exponents which modulate the importance
of $d_{sum}$ and $d_{min}$, respectively.
The values of $\beta_1$ and $\beta_2$ are generated using a multi-armed bandit approach \cite{multi-armed-bandit}, specifically based on the Upper Confidence Bound (UCB) strategy.
Finally, a simple roulette wheel method selects, using the probability values $p(g)$, the gate from $\mathcal{C}$ which is then added to $s$.





\section{Empirical Evaluation}
    \label{sec:evaluation}

In this section, we present experimental results obtained by applying the \textit{DIRSH} algorithm, implemented in C++, to a set of benchmark instances of QRP.
The objective of the analysis is to evaluate the effectiveness of the approach in terms of circuit depth and number of SWAP gates.

\subsection{Setup}
    \label{subsec:setup}


For our experimental evaluation, we employed the RevLib dataset~\cite{RevLib}, an online repository developed by the University of Bremen. RevLib provides a comprehensive collection of circuit specifications and benchmarks, widely used in the assessment of quantum synthesis, optimization, and compilation algorithms.




To assess the performance of \textit{DIRSH}, we compared it with the three variants of the \textit{LightSABRE} algorithm: \textit{basic}, \textit{lookahead}, and \textit{decay}.
Since \textit{DIRSH} is not able to search for a good initial qubit assignment, we impose that also \textit{LightSABRE} starts with the same
default qubit assignment.

The comparison was carried out on a subset of $150$ circuits of varying Size from the RevLib dataset, selected among those containing fewer than $100,000$ gates.
For a detailed description of each circuit instance (eg. depth, number of qubits and gate types) the reader can refer to Cowtan et al.\cite{QubitRouting}.
In our experiments, we use \textsc{ibmq\_tokyo} as the target quantum architecture, composed of $20$ qubits and $86$ connections.

\begin{table}[ht]
    \centering
    \renewcommand{\arraystretch}{1.4} 
    \begin{tabular}{|c|c|}
        \hline
        \textbf{Circuit Size $N$} & \textbf{Number of Chunks $n_c$} \\
        \hline
        $<\ 500$            & 1   \\
        $[500,\ 1000)$      & 10  \\
        $[1000,\ 10000)$    & 20  \\
        $[10000,\ 20000)$   & 50  \\
        $[20000,\ 50000)$   & 100 \\
        $\geq\ 50000$       & 150 \\
        \hline
    \end{tabular}
    \caption{Chunk division based on the circuit Size $N$.}
    \label{tab:scheme}
\end{table}

Each algorithm was tested under four different time budgets $T \in \{10, 20, 30, 60\}$ seconds.
For each run, we recorded the best solution obtained according to the two target objective functions, namely the \textit{depth} ($de$) and the \textit{number of SWAP gates} ($sw$) related to the obtained solution.
To account for stochastic variability, each configuration was executed 10 times, using 10 different random seeds.

Regarding the internal configuration of \textit{DIRSH}, the number of chunks ($n_c$) was instead chosen based on the number of gates ($N$) in the circuit, according to the scheme shown in Table~\ref{tab:scheme}.
This chunking strategy allowed us to tailor the algorithm's behavior to the Size of the input circuit.
It should be underlined that the maximum time allotted for the optimization of each chunk is $T/n_c$, in order to maintain the total runtime constant for each optimization run.


\subsection{Results}
\label{subsec:results}

\begin{figure}[H]     
    \centering
    \includegraphics[width=0.63\linewidth]{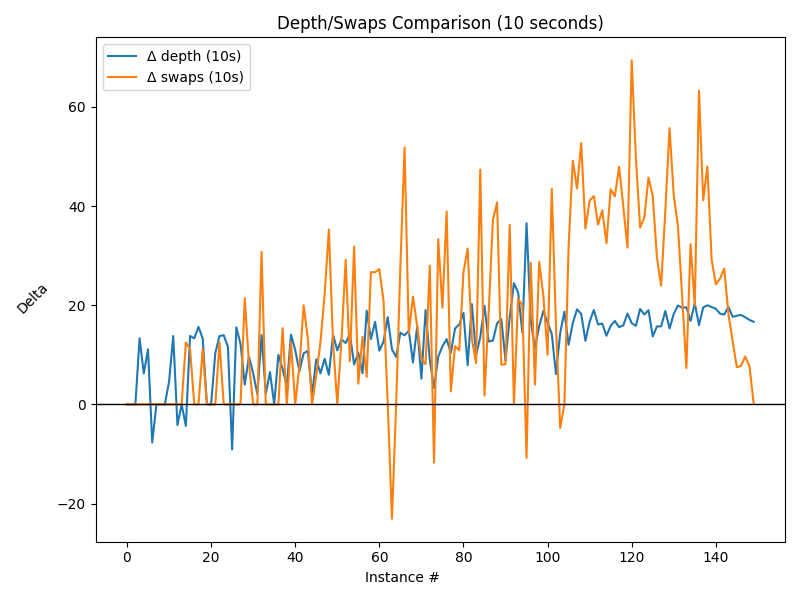}
    \caption{Values of $\Delta$ for Depth and SWAP with time budget $T = 10$s. Positive values indicate that DIRSH outperforms LightSABRE.}

    \label{fig:10_second_result}
\end{figure}

\begin{figure}[H]
    \centering
    \includegraphics[width=0.7\linewidth]{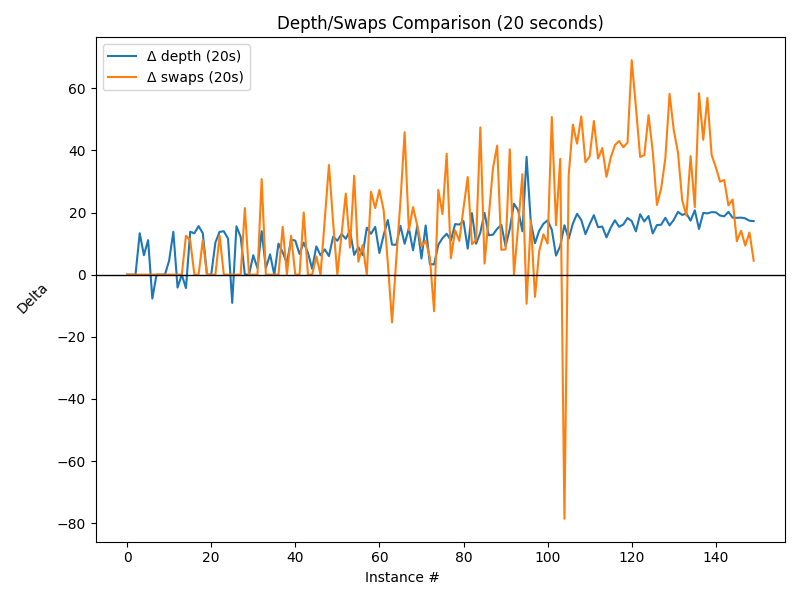}
    \caption{Values of $\Delta$ for Depth and SWAP with time budget $T = 20$s. Positive values indicate that DIRSH outperforms LightSABRE.}
    \label{fig:20_second_result}
\end{figure}

\begin{figure}[ht]
    \centering
    \includegraphics[width=0.7\linewidth]{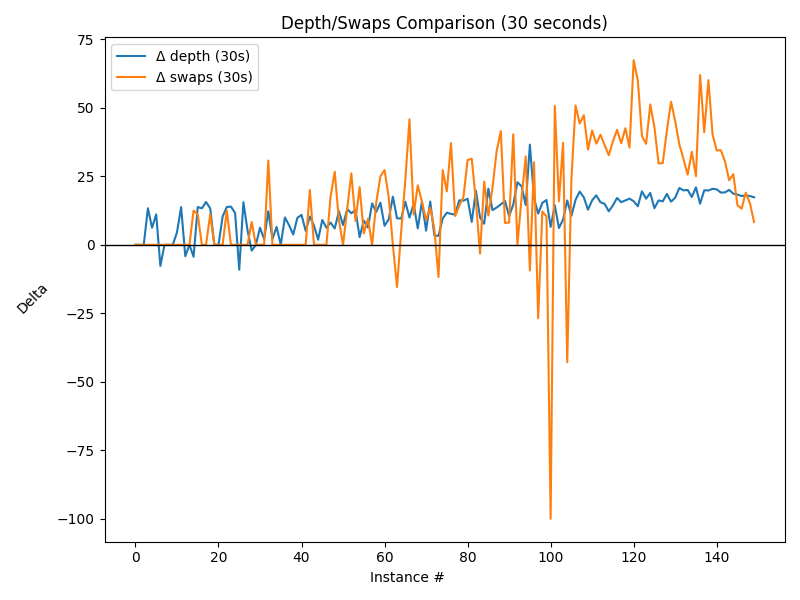}
    \caption{Values of $\Delta$ for Depth and SWAP with time budget $T = 30$s. Positive values indicate that DIRSH outperforms LightSABRE.}
    \label{fig:30_second_result}
\end{figure}

\begin{figure}[ht]
    \centering
    \includegraphics[width=0.7\linewidth]{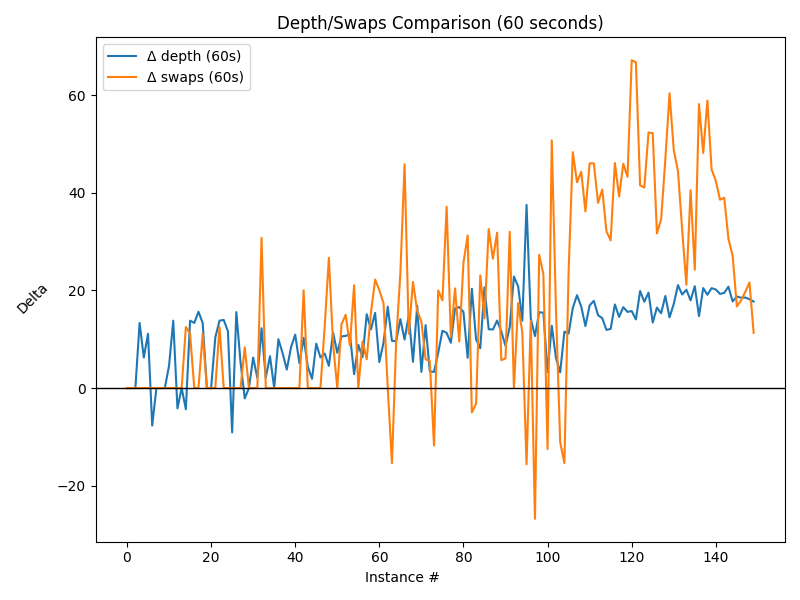}
    \caption{Values of $\Delta$ for Depth and SWAP with time budget $T = 60$s. Positive values indicate that DIRSH outperforms LightSABRE.}
    \label{fig:60_second_result}
\end{figure}

Figures~\ref{fig:10_second_result},~\ref{fig:20_second_result},~\ref{fig:30_second_result} and~\ref{fig:60_second_result} show the trend of the percentage deviation ($\Delta$) between the results obtained with \textit{LightSABRE} and \textit{DIRSH}, relatively to the two objectives of interest (circuit depth and number of SWAP gates) with time budget $T$ equal to $10$, $20$, $30$, and $60$ seconds, respectively. In each of the four figures, the x-axis shows the problem IDs, while the y-axis shows the percentage deviation computed between \textit{LightSABRE} and \textit{DIRSH}.
In more detail, given, for instance, the circuit depth $de$, the plotted percentage deviation is computed as follows (the same formula applies to $sw$): 

$$ \Delta = 100 \times \frac{de_{LightSabre} - de_{Dirsh}}{de_{LightSabre}} $$

From the very definition of the percentage deviation function, and considering both the functions of interest ($de$ and $sw$), it is clear that if the obtained values of $\Delta$ are greater than zero, then the \textit{DIRSH} algorithm is outperforming \textit{LightSABRE} (the higher the value, the better).  
In this respect, it is immediately observable from the figures that our algorithm performs better than \textit{LightSABRE} for the vast majority of problem instances, as all the plots are almost completely in the positive range for both $de$ and $sw$ functions.
Moreover, it can be interestingly observed from the figures that DIRSH's performance improvement exhibits an increasing trend with the circuit's Size.

\begin{table}[h]
     \centering
     \normalsize
     \caption{Comparison between \textit{DIRSH} and \textit{LightSABRE} ($D=Wins, -=Ties, L=Losses$).}
     \renewcommand{\arraystretch}{1.2}
     \begin{tabular}{c | c c c | c c c}
        \toprule
        & \multicolumn{3}{c|}{Swap minimization} & \multicolumn{3}{c}
        {Depth minimization} \\
        \toprule
        $T (secs)$ & D & - & L & D & - & L \\
        \midrule
        10 & 108 & 38 & 4 & 136 & 10 & 4 \\
        20 & 105 & 40 & 5 & 134 & 12 & 4 \\
        30 &  99 & 44 & 7 & 134 & 11 & 5 \\
        60 &  97 & 44 & 9 & 134 & 11 & 5 \\
        \bottomrule
    \end{tabular}
     \label{tab:comparison}
 \end{table}

Table~\ref{tab:comparison} conveys quantitative information about the number of times that \textit{DIRSH} outperforms \textit{LightSABRE}.
As the table shows, \textit{DIRSH} achieves rather good results: relatively to the $sw$ objective (\textit{LightSABRE}'s speciality), its performance never gets below $64.6\%$ of victories, $29.3\%$ of ties, and only $6.0\%$ of losses, while relatively to the $de$ objective, our performance never gets below $89.3\%$ of victories, $7.3\%$ of ties, and only $3.3\%$ of losses.

\begin{figure}[h]
    \centering
    \includegraphics[width=0.7\linewidth]{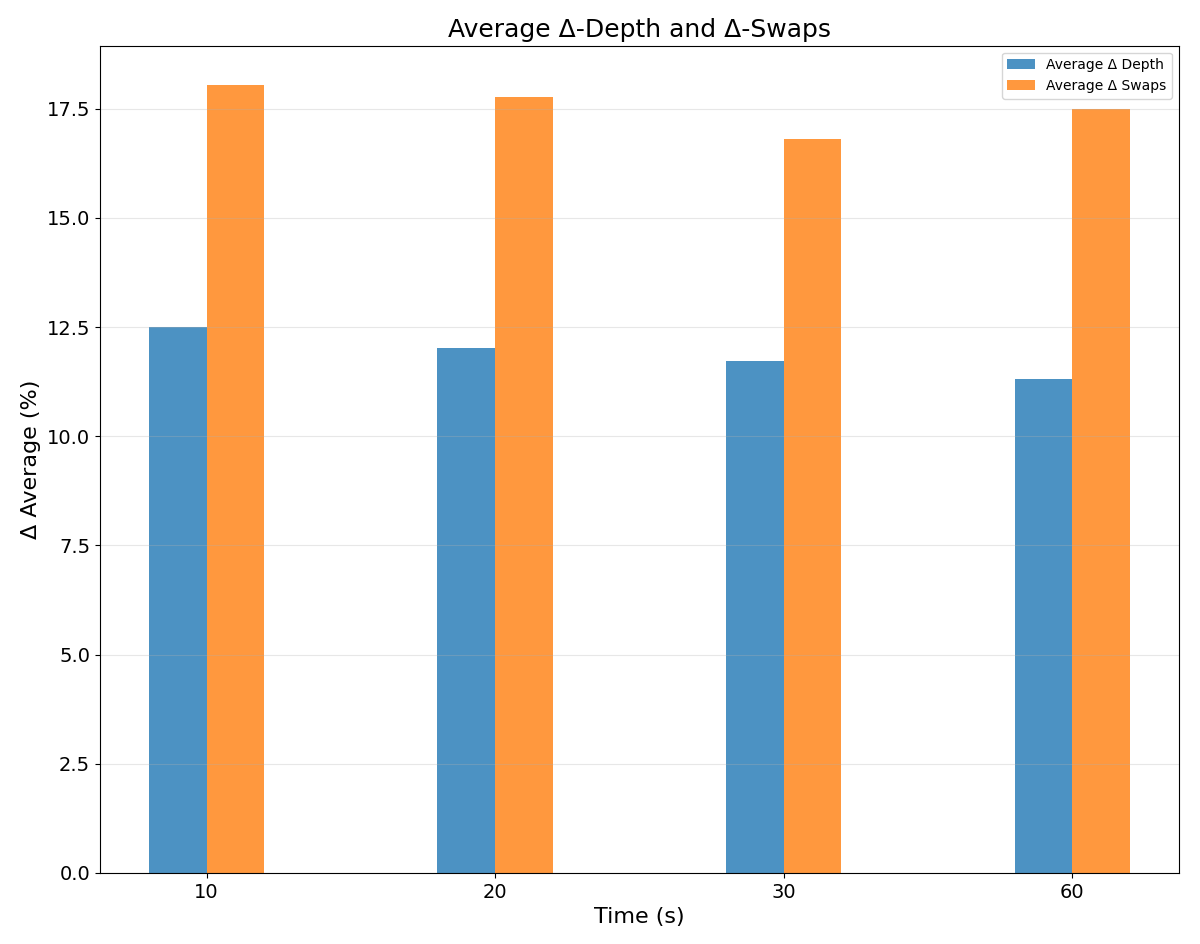}
    \caption{Average of the percentage deviations across all instances, in the depth minimization case (blue) and in the swap minimization case (orange).}
    \label{fig:analisys-fig}
\end{figure}

In addition, Figure~\ref{fig:analisys-fig} presents a cumulative view of both algorithms' performance in relation to the maximum CPU time allotted for resolution, by providing the average of the percentage deviations across all instances.
The constant positive values exhibited by the histogram confirms \textit{DIRSH}'s superiority over \textit{LightSABRE} for all time budgets.
For more detailed results, please refer to the tables \ref{tab:10s}, \ref{tab:20s}, \ref{tab:30s}, \ref{tab:60s} in the appendix.

\section{Conclusions and Future Work}
    \label{sec:conclusions}
    

This paper presents the \textit{DIRSH} algorithm, a heuristic-driven randomized, divide-et-impera search strategy for the qubit routing problem.
The algorithm divides the input quantum circuit in chunks, and optimizes each chunk exploiting a randomized heuristic selection of gates enhanced with a multi-armed bandit scheme, such that it balances global search, via random restarts and bandit sampling of $\beta_1,\beta_2$ parameters, with depth aware local pruning. 

Tested on 150 circuits of the RevLib benchmark mapped to the 20 qubit IBMQ Tokyo topology, \textit{DIRSH} consistently outperformed the three \textit{LightSABRE} variants in four time budgets ($10, 20, 30, 60s$).
In particular, with a budget of $60s$, it produced shorter circuits in $89.3\%$ of the instances and required fewer swaps in $64.6\%$ of them, winning $134/150$ the comparison on depth and $97/150$ the comparison on swaps.

These results confirm that the coupling between chunk-based decomposition and bandit-controlled adaptive heuristics is an effective recipe for routing generic quantum circuits on NISQ hardware.

For future developments, it is possible to devise an additional strategy that searches for a good initial qubit assignment.
Another possible development is to make \textit{DIRSH} noise-aware, by integrating the quantum gate error rates and measurements into the cost function, so that the algorithm directly reduces the overall noise during routing and produces more accurate circuits on real NISQ devices.
In addition, a further improvement of the \textit{DIRSH} algorithm could be based on the integration of a dynamic chunk splitting system, so that the static block Sizes (used in Table~\ref{tab:scheme}) would be computed on-line, which may help the algorithm converge faster on large instances.

\newpage

\appendix

\section{Additional Material}
    \label{sec:additional-material}

\scriptsize
\renewcommand{\arraystretch}{0.8}
\setlength{\tabcolsep}{13pt} 
\begin{center}
\begin{longtable}{lccccccc}
\caption{DIRSH VS SABRE 10 seconds}
\label{tab:10s} \\
\toprule
\textbf{Instance Name} & \textbf{Size} & \textbf{Sabre Depth} & \textbf{DIRSH Depth} & \textbf{Sabre Swaps} & \textbf{DIRSH Swaps} & \textbf{$\Delta$ Swap} & \textbf{$\Delta$ Depth} \\
\midrule
\endfirsthead

\toprule
\textbf{Instance Name} & \textbf{Size} & \textbf{Sabre Depth} & \textbf{DIRSH Depth} & \textbf{Sabre Swaps} & \textbf{DIRSH Swaps} & \textbf{$\Delta$ Swap} & \textbf{$\Delta$ Depth} \\
\midrule
\endhead

\bottomrule
\endfoot

\bottomrule
\endlastfoot
$\mathsf{graycode6\_47}$ & 5 & 5 & 5 & 3 & 3 & 0.00 & 0.00 \\
$\mathsf{ex1\_226}$ & 7 & 7 & 7 & 5 & 5 & 0.00 & 0.00 \\
$\mathsf{xor5\_254}$ & 7 & 7 & 7 & 5 & 5 & 0.00 & 0.00 \\
$\mathsf{4gt11\_84}$ & 18 & 15 & 13 & 4 & 4 & 0.00 & 13.33 \\
$\mathsf{ex-1\_166}$ & 19 & 16 & 15 & 3 & 3 & 0.00 & 6.25 \\
$\mathsf{ham3\_102}$ & 20 & 18 & 16 & 4 & 4 & 0.00 & 11.11 \\
$\mathsf{4mod5-v0\_20}$ & 20 & 13 & 14 & 4 & 4 & 0.00 & -7.69 \\
$\mathsf{4mod5-v1\_22}$ & 21 & 15 & 15 & 5 & 5 & 0.00 & 0.00 \\
$\mathsf{mod5d1\_63}$ & 22 & 16 & 16 & 6 & 6 & 0.00 & 0.00 \\
$\mathsf{4gt11\_83}$ & 23 & 16 & 16 & 5 & 5 & 0.00 & 0.00 \\
$\mathsf{4gt11\_82}$ & 27 & 22 & 21 & 7 & 7 & 0.00 & 4.55 \\
$\mathsf{rd32-v0\_66}$ & 34 & 29 & 25 & 8 & 8 & 0.00 & 13.79 \\
$\mathsf{4mod5-v0\_19}$ & 35 & 24 & 25 & 5 & 5 & 0.00 & -4.17 \\
$\mathsf{mod5mils\_65}$ & 35 & 24 & 24 & 6 & 6 & 0.00 & 0.00 \\
$\mathsf{4mod5-v1\_24}$ & 36 & 23 & 24 & 8 & 7 & 12.50 & -4.35 \\
$\mathsf{alu-v0\_27}$ & 36 & 29 & 25 & 9 & 8 & 11.11 & 13.79 \\
$\mathsf{rd32-v1\_68}$ & 36 & 30 & 26 & 8 & 8 & 0.00 & 13.33 \\
$\mathsf{3\_17\_13}$ & 36 & 32 & 27 & 7 & 7 & 0.00 & 15.62 \\
$\mathsf{alu-v4\_37}$ & 37 & 30 & 26 & 9 & 8 & 11.11 & 13.33 \\
$\mathsf{alu-v1\_28}$ & 37 & 26 & 26 & 7 & 7 & 0.00 & 0.00 \\
$\mathsf{alu-v3\_35}$ & 37 & 23 & 23 & 6 & 6 & 0.00 & 0.00 \\
$\mathsf{alu-v2\_33}$ & 37 & 29 & 26 & 9 & 9 & 0.00 & 10.34 \\
$\mathsf{alu-v1\_29}$ & 37 & 29 & 25 & 8 & 7 & 12.50 & 13.79 \\
$\mathsf{miller\_11}$ & 50 & 43 & 37 & 9 & 9 & 0.00 & 13.95 \\
$\mathsf{decod24-v0\_38}$ & 51 & 43 & 38 & 11 & 11 & 0.00 & 11.63 \\
$\mathsf{alu-v3\_34}$ & 52 & 33 & 36 & 7 & 7 & 0.00 & -9.09 \\
$\mathsf{decod24-v2\_43}$ & 52 & 45 & 38 & 11 & 11 & 0.00 & 15.56 \\
$\mathsf{mod5d2\_64}$ & 53 & 41 & 36 & 11 & 11 & 0.00 & 12.20 \\
$\mathsf{4gt13\_92}$ & 66 & 50 & 48 & 14 & 11 & 21.43 & 4.00 \\
$\mathsf{4gt13-v1\_93}$ & 68 & 52 & 47 & 12 & 11 & 8.33 & 9.62 \\
$\mathsf{4mod5-v0\_18}$ & 69 & 48 & 45 & 12 & 12 & 0.00 & 6.25 \\
$\mathsf{4mod5-v1\_23}$ & 69 & 48 & 47 & 12 & 12 & 0.00 & 2.08 \\
$\mathsf{one-two-three-v2\_100}$ & 69 & 50 & 43 & 13 & 9 & 30.77 & 14.00 \\
$\mathsf{one-two-three-v3\_101}$ & 70 & 47 & 46 & 11 & 11 & 0.00 & 2.13 \\
$\mathsf{4mod5-bdd\_287}$ & 70 & 46 & 43 & 8 & 8 & 0.00 & 6.52 \\
$\mathsf{qe\_qft\_4}$ & 71 & 39 & 39 & 5 & 5 & 0.00 & 0.00 \\
$\mathsf{decod24-bdd\_294}$ & 73 & 50 & 45 & 10 & 10 & 0.00 & 10.00 \\
$\mathsf{4gt5\_75}$ & 83 & 56 & 52 & 13 & 11 & 15.38 & 7.14 \\
$\mathsf{alu-bdd\_288}$ & 84 & 53 & 51 & 10 & 10 & 0.00 & 3.77 \\
$\mathsf{rd32\_270}$ & 84 & 64 & 55 & 16 & 14 & 12.50 & 14.06 \\
$\mathsf{alu-v0\_26}$ & 84 & 64 & 57 & 14 & 14 & 0.00 & 10.94 \\
$\mathsf{decod24-v1\_41}$ & 85 & 60 & 56 & 13 & 12 & 7.69 & 6.67 \\
$\mathsf{4gt5\_76}$ & 91 & 68 & 61 & 15 & 12 & 20.00 & 10.29 \\
$\mathsf{4gt13\_91}$ & 103 & 74 & 66 & 15 & 13 & 13.33 & 10.81 \\
$\mathsf{qe\_qft\_5}$ & 107 & 53 & 52 & 2 & 2 & 0.00 & 1.89 \\
$\mathsf{4gt13\_90}$ & 107 & 77 & 70 & 17 & 16 & 5.88 & 9.09 \\
$\mathsf{alu-v4\_36}$ & 115 & 80 & 75 & 16 & 14 & 12.50 & 6.25 \\
$\mathsf{4gt5\_77}$ & 131 & 87 & 79 & 18 & 14 & 22.22 & 9.20 \\
$\mathsf{rd53\_138}$ & 132 & 67 & 63 & 17 & 11 & 35.29 & 5.97 \\
$\mathsf{one-two-three-v1\_99}$ & 132 & 100 & 86 & 24 & 21 & 12.50 & 14.00 \\
$\mathsf{one-two-three-v0\_98}$ & 146 & 101 & 90 & 19 & 19 & 0.00 & 10.89 \\
$\mathsf{4gt10-v1\_81}$ & 148 & 107 & 93 & 23 & 20 & 13.04 & 13.08 \\
$\mathsf{decod24-v3\_45}$ & 150 & 105 & 92 & 24 & 17 & 29.17 & 12.38 \\
$\mathsf{aj-e11\_165}$ & 151 & 113 & 97 & 23 & 21 & 8.70 & 14.16 \\
$\mathsf{4mod7-v0\_94}$ & 162 & 112 & 103 & 22 & 15 & 31.82 & 8.04 \\
$\mathsf{alu-v2\_32}$ & 163 & 115 & 103 & 24 & 23 & 4.17 & 10.43 \\
$\mathsf{4mod7-v1\_96}$ & 164 & 111 & 104 & 22 & 19 & 13.64 & 6.31 \\
$\mathsf{mini\_alu\_305}$ & 173 & 90 & 73 & 18 & 17 & 5.56 & 18.89 \\
$\mathsf{cnt3-5\_179}$ & 175 & 76 & 66 & 30 & 22 & 26.67 & 13.16 \\
$\mathsf{mod10\_176}$ & 178 & 132 & 110 & 30 & 22 & 26.67 & 16.67 \\
$\mathsf{4gt4-v0\_80}$ & 179 & 120 & 107 & 22 & 16 & 27.27 & 10.83 \\
$\mathsf{4gt12-v0\_88}$ & 194 & 134 & 117 & 24 & 19 & 20.83 & 12.69 \\
$\mathsf{qft\_10}$ & 200 & 91 & 75 & 26 & 26 & 0.00 & 17.58 \\
$\mathsf{0410184\_169}$ & 211 & 126 & 112 & 26 & 32 & -23.08 & 11.11 \\
$\mathsf{sys6-v0\_111}$ & 215 & 94 & 85 & 21 & 21 & 0.00 & 9.57 \\
$\mathsf{4\_49\_16}$ & 217 & 159 & 136 & 36 & 26 & 27.78 & 14.47 \\
$\mathsf{4gt12-v1\_89}$ & 228 & 158 & 136 & 27 & 13 & 51.85 & 13.92 \\
$\mathsf{rd73\_140}$ & 230 & 121 & 103 & 28 & 24 & 14.29 & 14.88 \\
$\mathsf{4gt4-v0\_79}$ & 231 & 154 & 141 & 23 & 18 & 21.74 & 8.44 \\
$\mathsf{hwb4\_49}$ & 233 & 178 & 150 & 38 & 32 & 15.79 & 15.73 \\
$\mathsf{4gt4-v0\_78}$ & 235 & 154 & 146 & 22 & 20 & 9.09 & 5.19 \\
$\mathsf{mod10\_171}$ & 244 & 184 & 149 & 37 & 34 & 8.11 & 19.02 \\
$\mathsf{4gt12-v0\_87}$ & 247 & 156 & 142 & 25 & 18 & 28.00 & 8.97 \\
$\mathsf{4gt12-v0\_86}$ & 251 & 152 & 147 & 17 & 19 & -11.76 & 3.29 \\
$\mathsf{4gt4-v0\_72}$ & 258 & 156 & 141 & 24 & 16 & 33.33 & 9.62 \\
$\mathsf{sym6\_316}$ & 270 & 162 & 143 & 41 & 33 & 19.51 & 11.73 \\
$\mathsf{4gt4-v1\_74}$ & 273 & 190 & 165 & 36 & 22 & 38.89 & 13.16 \\
$\mathsf{rd53\_311}$ & 275 & 154 & 138 & 38 & 37 & 2.63 & 10.39 \\
$\mathsf{mini-alu\_167}$ & 288 & 209 & 177 & 51 & 45 & 11.76 & 15.31 \\
$\mathsf{one-two-three-v0\_97}$ & 290 & 211 & 177 & 46 & 41 & 10.87 & 16.11 \\
$\mathsf{rd53\_135}$ & 296 & 206 & 168 & 45 & 33 & 26.67 & 18.45 \\
$\mathsf{ham7\_104}$ & 320 & 215 & 198 & 35 & 24 & 31.43 & 7.91 \\
$\mathsf{sym9\_146}$ & 328 & 183 & 146 & 53 & 46 & 13.21 & 20.22 \\
$\mathsf{decod24-enable\_126}$ & 338 & 221 & 199 & 36 & 33 & 8.33 & 9.95 \\
$\mathsf{mod8-10\_178}$ & 342 & 236 & 204 & 38 & 20 & 47.37 & 13.56 \\
$\mathsf{rd84\_142}$ & 343 & 161 & 129 & 56 & 55 & 1.79 & 19.88 \\
$\mathsf{4gt4-v0\_73}$ & 395 & 276 & 241 & 43 & 35 & 18.60 & 12.68 \\
$\mathsf{ex3\_229}$ & 403 & 272 & 237 & 43 & 27 & 37.21 & 12.87 \\
$\mathsf{mod8-10\_177}$ & 440 & 318 & 266 & 54 & 32 & 40.74 & 16.35 \\
$\mathsf{alu-v2\_31}$ & 451 & 343 & 284 & 75 & 69 & 8.00 & 17.20 \\
$\mathsf{C17\_204}$ & 467 & 295 & 269 & 37 & 34 & 8.11 & 8.81 \\
$\mathsf{rd53\_131}$ & 469 & 331 & 273 & 58 & 37 & 36.21 & 17.52 \\
$\mathsf{ising\_model\_10}$ & 480 & 94 & 71 & 6 & 6 & 0.00 & 24.47 \\
$\mathsf{cnt3-5\_180}$ & 485 & 297 & 230 & 75 & 59 & 21.33 & 22.56 \\
$\mathsf{alu-v2\_30}$ & 504 & 358 & 306 & 65 & 52 & 20.00 & 14.53 \\
$\mathsf{qft\_16}$ & 512 & 219 & 139 & 65 & 72 & -10.77 & 36.53 \\
$\mathsf{mod5adder\_127}$ & 555 & 384 & 319 & 63 & 45 & 28.57 & 16.93 \\
$\mathsf{rd53\_133}$ & 580 & 399 & 354 & 75 & 72 & 4.00 & 11.28 \\
$\mathsf{majority\_239}$ & 612 & 425 & 358 & 66 & 47 & 28.79 & 15.76 \\
$\mathsf{ex2\_227}$ & 631 & 461 & 374 & 87 & 68 & 21.84 & 18.87 \\
$\mathsf{ising\_model\_13}$ & 633 & 103 & 86 & 10 & 9 & 10.00 & 16.50 \\
$\mathsf{cm82a\_208}$ & 650 & 416 & 357 & 69 & 39 & 43.48 & 14.18 \\
$\mathsf{sf\_276}$ & 778 & 490 & 460 & 44 & 37 & 15.91 & 6.12 \\
$\mathsf{sf\_274}$ & 781 & 528 & 452 & 63 & 66 & -4.76 & 14.39 \\
$\mathsf{ising\_model\_16}$ & 786 & 107 & 87 & 15 & 15 & 0.00 & 18.69 \\
$\mathsf{con1\_216}$ & 954 & 607 & 534 & 88 & 60 & 31.82 & 12.03 \\
$\mathsf{wim\_266}$ & 986 & 646 & 540 & 118 & 60 & 49.15 & 16.41 \\
$\mathsf{rd53\_130}$ & 1043 & 751 & 607 & 140 & 79 & 43.57 & 19.17 \\
$\mathsf{f2\_232}$ & 1206 & 867 & 709 & 167 & 79 & 52.69 & 18.22 \\
$\mathsf{cm152a\_212}$ & 1221 & 827 & 721 & 141 & 91 & 35.46 & 12.82 \\
$\mathsf{rd53\_251}$ & 1291 & 911 & 759 & 163 & 96 & 41.10 & 16.68 \\
$\mathsf{hwb5\_53}$ & 1336 & 987 & 799 & 188 & 109 & 42.02 & 19.05 \\
$\mathsf{pm1\_249}$ & 1776 & 1172 & 983 & 193 & 123 & 36.27 & 16.13 \\
$\mathsf{cm42a\_207}$ & 1776 & 1178 & 986 & 189 & 115 & 39.15 & 16.30 \\
$\mathsf{dc1\_220}$ & 1914 & 1257 & 1083 & 166 & 112 & 32.53 & 13.84 \\
$\mathsf{squar5\_261}$ & 1993 & 1313 & 1105 & 205 & 116 & 43.41 & 15.84 \\
$\mathsf{sqrt8\_260}$ & 3009 & 2104 & 1750 & 379 & 220 & 41.95 & 16.83 \\
$\mathsf{z4\_268}$ & 3073 & 2042 & 1724 & 340 & 177 & 47.94 & 15.57 \\
$\mathsf{radd\_250}$ & 3213 & 2223 & 1869 & 383 & 230 & 39.95 & 15.92 \\
$\mathsf{adr4\_197}$ & 3439 & 2362 & 1929 & 386 & 264 & 31.61 & 18.33 \\
$\mathsf{sym6\_145}$ & 3888 & 2766 & 2313 & 461 & 141 & 69.41 & 16.38 \\
$\mathsf{misex1\_241}$ & 4813 & 3274 & 2755 & 477 & 240 & 49.69 & 15.85 \\
$\mathsf{rd73\_252}$ & 5321 & 3756 & 3033 & 684 & 440 & 35.67 & 19.25 \\
$\mathsf{cycle10\_2\_110}$ & 6050 & 4408 & 3609 & 795 & 495 & 37.74 & 18.13 \\
$\mathsf{hwb6\_56}$ & 6723 & 4928 & 3991 & 918 & 498 & 45.75 & 19.01 \\
$\mathsf{square\_root\_7}$ & 7630 & 4719 & 4072 & 734 & 425 & 42.10 & 13.71 \\
$\mathsf{ham15\_107}$ & 8763 & 6126 & 5163 & 1072 & 755 & 29.57 & 15.72 \\
$\mathsf{dc2\_222}$ & 9462 & 6672 & 5621 & 1156 & 879 & 23.96 & 15.75 \\
$\mathsf{sqn\_258}$ & 10223 & 7108 & 5770 & 1297 & 790 & 39.09 & 18.82 \\
$\mathsf{inc\_237}$ & 10619 & 7232 & 6125 & 1127 & 499 & 55.72 & 15.31 \\
$\mathsf{cm85a\_209}$ & 11414 & 8144 & 6658 & 1447 & 837 & 42.16 & 18.25 \\
$\mathsf{rd84\_253}$ & 13658 & 9773 & 7824 & 2056 & 1317 & 35.94 & 19.94 \\
$\mathsf{root\_255}$ & 17159 & 11970 & 9637 & 2398 & 1877 & 21.73 & 19.49 \\
$\mathsf{co14\_215}$ & 17936 & 11719 & 9424 & 2535 & 2349 & 7.34 & 19.58 \\
$\mathsf{mlp4\_245}$ & 18852 & 13376 & 11125 & 2474 & 1675 & 32.30 & 16.83 \\
$\mathsf{urf2\_277}$ & 20112 & 15508 & 12343 & 3385 & 2712 & 19.88 & 20.41 \\
$\mathsf{sym9\_148}$ & 21504 & 14851 & 12479 & 2297 & 843 & 63.30 & 15.97 \\
$\mathsf{life\_238}$ & 22445 & 16531 & 13297 & 3251 & 1912 & 41.19 & 19.56 \\
$\mathsf{hwb7\_59}$ & 24379 & 17821 & 14258 & 3515 & 1828 & 47.99 & 19.99 \\
$\mathsf{max46\_240}$ & 27126 & 19153 & 15391 & 3807 & 2693 & 29.26 & 19.64 \\
$\mathsf{clip\_206}$ & 33827 & 24021 & 19388 & 4725 & 3581 & 24.21 & 19.29 \\
$\mathsf{sym9\_193}$ & 34881 & 25642 & 20948 & 5123 & 3824 & 25.36 & 18.31 \\
$\mathsf{9symml\_195}$ & 34881 & 25581 & 20945 & 5094 & 3699 & 27.39 & 18.12 \\
$\mathsf{dist\_223}$ & 38046 & 27204 & 21850 & 5646 & 4615 & 18.26 & 19.68 \\
$\mathsf{sao2\_257}$ & 38577 & 26214 & 21584 & 5132 & 4479 & 12.72 & 17.66 \\
$\mathsf{urf5\_280}$ & 49829 & 38073 & 31263 & 8391 & 7763 & 7.48 & 17.89 \\
$\mathsf{urf1\_278}$ & 54766 & 42091 & 34496 & 9422 & 8687 & 7.80 & 18.04 \\
$\mathsf{sym10\_262}$ & 64283 & 48312 & 39824 & 10125 & 9147 & 9.66 & 17.57 \\
$\mathsf{hwb8\_113}$ & 69380 & 51750 & 42930 & 10664 & 9846 & 7.67 & 17.04 \\
$\mathsf{urf2\_152}$ & 80480 & 58660 & 48884 & 11805 & 11780 & 0.21 & 16.67 \\
\end{longtable}
\end{center}

\scriptsize
\renewcommand{\arraystretch}{0.8}
\setlength{\tabcolsep}{13pt} 
\begin{center}
\begin{longtable}{lccccccc}
\caption{DIRSH VS SABRE 20 seconds}
\label{tab:20s} \\
\toprule
\textbf{Instance Name} & \textbf{Size} & \textbf{Sabre Depth} & \textbf{DIRSH Depth} & \textbf{Sabre Swaps} & \textbf{DIRSH Swaps} & \textbf{$\Delta$ Swap} & \textbf{$\Delta$ Depth} \\
\midrule
\endfirsthead

\toprule
\textbf{Instance Name} & \textbf{Size} & \textbf{Sabre Depth} & \textbf{DIRSH Depth} & \textbf{Sabre Swaps} & \textbf{DIRSH Swaps} & \textbf{$\Delta$ Swap} & \textbf{$\Delta$ Depth} \\
\midrule
\endhead

\bottomrule
\endfoot

\bottomrule
\endlastfoot
$\mathsf{graycode6\_47}$ & 5 & 5 & 5 & 3 & 3 & 0.00 & 0.00 \\
$\mathsf{ex1\_226}$ & 7 & 7 & 7 & 5 & 5 & 0.00 & 0.00 \\
$\mathsf{xor5\_254}$ & 7 & 7 & 7 & 5 & 5 & 0.00 & 0.00 \\
$\mathsf{4gt11\_84}$ & 18 & 15 & 13 & 4 & 4 & 0.00 & 13.33 \\
$\mathsf{ex-1\_166}$ & 19 & 16 & 15 & 3 & 3 & 0.00 & 6.25 \\
$\mathsf{ham3\_102}$ & 20 & 18 & 16 & 4 & 4 & 0.00 & 11.11 \\
$\mathsf{4mod5-v0\_20}$ & 20 & 13 & 14 & 4 & 4 & 0.00 & -7.69 \\
$\mathsf{4mod5-v1\_22}$ & 21 & 15 & 15 & 5 & 5 & 0.00 & 0.00 \\
$\mathsf{mod5d1\_63}$ & 22 & 16 & 16 & 6 & 6 & 0.00 & 0.00 \\
$\mathsf{4gt11\_83}$ & 23 & 16 & 16 & 5 & 5 & 0.00 & 0.00 \\
$\mathsf{4gt11\_82}$ & 27 & 22 & 21 & 7 & 7 & 0.00 & 4.55 \\
$\mathsf{rd32-v0\_66}$ & 34 & 29 & 25 & 8 & 8 & 0.00 & 13.79 \\
$\mathsf{4mod5-v0\_19}$ & 35 & 24 & 25 & 5 & 5 & 0.00 & -4.17 \\
$\mathsf{mod5mils\_65}$ & 35 & 24 & 24 & 6 & 6 & 0.00 & 0.00 \\
$\mathsf{4mod5-v1\_24}$ & 36 & 23 & 24 & 8 & 7 & 12.50 & -4.35 \\
$\mathsf{alu-v0\_27}$ & 36 & 29 & 25 & 9 & 8 & 11.11 & 13.79 \\
$\mathsf{rd32-v1\_68}$ & 36 & 30 & 26 & 8 & 8 & 0.00 & 13.33 \\
$\mathsf{3\_17\_13}$ & 36 & 32 & 27 & 7 & 7 & 0.00 & 15.62 \\
$\mathsf{alu-v4\_37}$ & 37 & 30 & 26 & 9 & 8 & 11.11 & 13.33 \\
$\mathsf{alu-v1\_28}$ & 37 & 26 & 26 & 7 & 7 & 0.00 & 0.00 \\
$\mathsf{alu-v3\_35}$ & 37 & 23 & 23 & 6 & 6 & 0.00 & 0.00 \\
$\mathsf{alu-v2\_33}$ & 37 & 29 & 26 & 9 & 9 & 0.00 & 10.34 \\
$\mathsf{alu-v1\_29}$ & 37 & 29 & 25 & 8 & 7 & 12.50 & 13.79 \\
$\mathsf{miller\_11}$ & 50 & 43 & 37 & 9 & 9 & 0.00 & 13.95 \\
$\mathsf{decod24-v0\_38}$ & 51 & 43 & 38 & 11 & 11 & 0.00 & 11.63 \\
$\mathsf{alu-v3\_34}$ & 52 & 33 & 36 & 7 & 7 & 0.00 & -9.09 \\
$\mathsf{decod24-v2\_43}$ & 52 & 45 & 38 & 11 & 11 & 0.00 & 15.56 \\
$\mathsf{mod5d2\_64}$ & 53 & 41 & 36 & 11 & 11 & 0.00 & 12.20 \\
$\mathsf{4gt13\_92}$ & 66 & 48 & 48 & 14 & 11 & 21.43 & 0.00 \\
$\mathsf{4gt13-v1\_93}$ & 68 & 47 & 47 & 11 & 11 & 0.00 & 0.00 \\
$\mathsf{4mod5-v0\_18}$ & 69 & 48 & 45 & 12 & 12 & 0.00 & 6.25 \\
$\mathsf{4mod5-v1\_23}$ & 69 & 48 & 47 & 12 & 12 & 0.00 & 2.08 \\
$\mathsf{one-two-three-v2\_100}$ & 69 & 50 & 43 & 13 & 9 & 30.77 & 14.00 \\
$\mathsf{one-two-three-v3\_101}$ & 70 & 47 & 46 & 11 & 11 & 0.00 & 2.13 \\
$\mathsf{4mod5-bdd\_287}$ & 70 & 46 & 43 & 8 & 8 & 0.00 & 6.52 \\
$\mathsf{qe\_qft\_4}$ & 71 & 39 & 39 & 5 & 5 & 0.00 & 0.00 \\
$\mathsf{decod24-bdd\_294}$ & 73 & 50 & 45 & 10 & 10 & 0.00 & 10.00 \\
$\mathsf{4gt5\_75}$ & 83 & 56 & 52 & 13 & 11 & 15.38 & 7.14 \\
$\mathsf{alu-bdd\_288}$ & 84 & 53 & 51 & 10 & 10 & 0.00 & 3.77 \\
$\mathsf{rd32\_270}$ & 84 & 62 & 55 & 16 & 14 & 12.50 & 11.29 \\
$\mathsf{alu-v0\_26}$ & 84 & 64 & 57 & 14 & 14 & 0.00 & 10.94 \\
$\mathsf{decod24-v1\_41}$ & 85 & 60 & 56 & 12 & 12 & 0.00 & 6.67 \\
$\mathsf{4gt5\_76}$ & 91 & 68 & 61 & 15 & 12 & 20.00 & 10.29 \\
$\mathsf{4gt13\_91}$ & 103 & 71 & 66 & 13 & 13 & 0.00 & 7.04 \\
$\mathsf{qe\_qft\_5}$ & 107 & 53 & 52 & 2 & 2 & 0.00 & 1.89 \\
$\mathsf{4gt13\_90}$ & 107 & 77 & 70 & 17 & 16 & 5.88 & 9.09 \\
$\mathsf{alu-v4\_36}$ & 115 & 80 & 75 & 14 & 14 & 0.00 & 6.25 \\
$\mathsf{4gt5\_77}$ & 131 & 86 & 79 & 17 & 14 & 17.65 & 8.14 \\
$\mathsf{rd53\_138}$ & 132 & 67 & 63 & 17 & 11 & 35.29 & 5.97 \\
$\mathsf{one-two-three-v1\_99}$ & 132 & 98 & 86 & 24 & 20 & 16.67 & 12.24 \\
$\mathsf{one-two-three-v0\_98}$ & 146 & 101 & 90 & 19 & 19 & 0.00 & 10.89 \\
$\mathsf{4gt10-v1\_81}$ & 148 & 107 & 93 & 23 & 20 & 13.04 & 13.08 \\
$\mathsf{decod24-v3\_45}$ & 150 & 104 & 92 & 23 & 17 & 26.09 & 11.54 \\
$\mathsf{aj-e11\_165}$ & 151 & 113 & 97 & 23 & 21 & 8.70 & 14.16 \\
$\mathsf{4mod7-v0\_94}$ & 162 & 110 & 103 & 22 & 15 & 31.82 & 6.36 \\
$\mathsf{alu-v2\_32}$ & 163 & 113 & 103 & 24 & 23 & 4.17 & 8.85 \\
$\mathsf{4mod7-v1\_96}$ & 164 & 111 & 104 & 21 & 19 & 9.52 & 6.31 \\
$\mathsf{mini\_alu\_305}$ & 173 & 86 & 73 & 17 & 17 & 0.00 & 15.12 \\
$\mathsf{cnt3-5\_179}$ & 175 & 76 & 66 & 30 & 22 & 26.67 & 13.16 \\
$\mathsf{mod10\_176}$ & 178 & 130 & 110 & 28 & 22 & 21.43 & 15.38 \\
$\mathsf{4gt4-v0\_80}$ & 179 & 115 & 107 & 22 & 16 & 27.27 & 6.96 \\
$\mathsf{4gt12-v0\_88}$ & 194 & 134 & 117 & 24 & 19 & 20.83 & 12.69 \\
$\mathsf{qft\_10}$ & 200 & 91 & 75 & 26 & 25 & 3.85 & 17.58 \\
$\mathsf{0410184\_169}$ & 211 & 124 & 112 & 26 & 30 & -15.38 & 9.68 \\
$\mathsf{sys6-v0\_111}$ & 215 & 94 & 85 & 21 & 20 & 4.76 & 9.57 \\
$\mathsf{4\_49\_16}$ & 217 & 159 & 134 & 34 & 26 & 23.53 & 15.72 \\
$\mathsf{4gt12-v1\_89}$ & 228 & 151 & 136 & 24 & 13 & 45.83 & 9.93 \\
$\mathsf{rd73\_140}$ & 230 & 121 & 103 & 28 & 24 & 14.29 & 14.88 \\
$\mathsf{4gt4-v0\_79}$ & 231 & 153 & 141 & 23 & 18 & 21.74 & 7.84 \\
$\mathsf{hwb4\_49}$ & 233 & 178 & 150 & 38 & 32 & 15.79 & 15.73 \\
$\mathsf{4gt4-v0\_78}$ & 235 & 154 & 146 & 22 & 20 & 9.09 & 5.19 \\
$\mathsf{mod10\_171}$ & 244 & 177 & 149 & 37 & 33 & 10.81 & 15.82 \\
$\mathsf{4gt12-v0\_87}$ & 247 & 147 & 142 & 18 & 17 & 5.56 & 3.40 \\
$\mathsf{4gt12-v0\_86}$ & 251 & 152 & 147 & 17 & 19 & -11.76 & 3.29 \\
$\mathsf{4gt4-v0\_72}$ & 258 & 156 & 141 & 22 & 16 & 27.27 & 9.62 \\
$\mathsf{sym6\_316}$ & 270 & 162 & 143 & 41 & 33 & 19.51 & 11.73 \\
$\mathsf{4gt4-v1\_74}$ & 273 & 190 & 165 & 36 & 22 & 38.89 & 13.16 \\
$\mathsf{rd53\_311}$ & 275 & 154 & 137 & 38 & 36 & 5.26 & 11.04 \\
$\mathsf{mini-alu\_167}$ & 288 & 209 & 175 & 49 & 42 & 14.29 & 16.27 \\
$\mathsf{one-two-three-v0\_97}$ & 290 & 211 & 177 & 46 & 41 & 10.87 & 16.11 \\
$\mathsf{rd53\_135}$ & 296 & 203 & 168 & 42 & 33 & 21.43 & 17.24 \\
$\mathsf{ham7\_104}$ & 320 & 215 & 197 & 35 & 24 & 31.43 & 8.37 \\
$\mathsf{sym9\_146}$ & 328 & 182 & 146 & 51 & 46 & 9.80 & 19.78 \\
$\mathsf{decod24-enable\_126}$ & 338 & 221 & 199 & 36 & 32 & 11.11 & 9.95 \\
$\mathsf{mod8-10\_178}$ & 342 & 236 & 204 & 38 & 20 & 47.37 & 13.56 \\
$\mathsf{rd84\_142}$ & 343 & 161 & 129 & 56 & 54 & 3.57 & 19.88 \\
$\mathsf{4gt4-v0\_73}$ & 395 & 276 & 241 & 43 & 35 & 18.60 & 12.68 \\
$\mathsf{ex3\_229}$ & 403 & 272 & 237 & 38 & 25 & 34.21 & 12.87 \\
$\mathsf{mod8-10\_177}$ & 440 & 312 & 266 & 53 & 31 & 41.51 & 14.74 \\
$\mathsf{alu-v2\_31}$ & 451 & 337 & 283 & 75 & 69 & 8.00 & 16.02 \\
$\mathsf{C17\_204}$ & 467 & 295 & 268 & 37 & 34 & 8.11 & 9.15 \\
$\mathsf{rd53\_131}$ & 469 & 320 & 273 & 57 & 34 & 40.35 & 14.69 \\
$\mathsf{ising\_model\_10}$ & 480 & 92 & 71 & 6 & 6 & 0.00 & 22.83 \\
$\mathsf{cnt3-5\_180}$ & 485 & 290 & 230 & 70 & 59 & 15.71 & 20.69 \\
$\mathsf{alu-v2\_30}$ & 504 & 358 & 308 & 65 & 44 & 32.31 & 13.97 \\
$\mathsf{qft\_16}$ & 512 & 219 & 136 & 64 & 70 & -9.38 & 37.90 \\
$\mathsf{mod5adder\_127}$ & 555 & 384 & 319 & 63 & 52 & 17.46 & 16.93 \\
$\mathsf{rd53\_133}$ & 580 & 395 & 355 & 56 & 60 & -7.14 & 10.13 \\
$\mathsf{majority\_239}$ & 612 & 425 & 365 & 66 & 61 & 7.58 & 14.12 \\
$\mathsf{ex2\_227}$ & 631 & 446 & 373 & 77 & 67 & 12.99 & 16.37 \\
$\mathsf{ising\_model\_13}$ & 633 & 103 & 85 & 10 & 9 & 10.00 & 17.48 \\
$\mathsf{cm82a\_208}$ & 650 & 416 & 356 & 69 & 34 & 50.72 & 14.42 \\
$\mathsf{sf\_276}$ & 778 & 490 & 460 & 44 & 37 & 15.91 & 6.12 \\
$\mathsf{sf\_274}$ & 781 & 498 & 452 & 51 & 32 & 37.25 & 9.24 \\
$\mathsf{ising\_model\_16}$ & 786 & 107 & 90 & 14 & 25 & -78.57 & 15.89 \\
$\mathsf{con1\_216}$ & 954 & 607 & 537 & 88 & 60 & 31.82 & 11.53 \\
$\mathsf{wim\_266}$ & 986 & 646 & 540 & 118 & 61 & 48.31 & 16.41 \\
$\mathsf{rd53\_130}$ & 1043 & 751 & 604 & 140 & 81 & 42.14 & 19.57 \\
$\mathsf{f2\_232}$ & 1206 & 863 & 711 & 167 & 82 & 50.90 & 17.61 \\
$\mathsf{cm152a\_212}$ & 1221 & 827 & 719 & 141 & 90 & 36.17 & 13.06 \\
$\mathsf{rd53\_251}$ & 1291 & 911 & 763 & 163 & 101 & 38.04 & 16.25 \\
$\mathsf{hwb5\_53}$ & 1336 & 987 & 798 & 188 & 95 & 49.47 & 19.15 \\
$\mathsf{pm1\_249}$ & 1776 & 1166 & 988 & 190 & 119 & 37.37 & 15.27 \\
$\mathsf{cm42a\_207}$ & 1776 & 1176 & 994 & 189 & 112 & 40.74 & 15.48 \\
$\mathsf{dc1\_220}$ & 1914 & 1235 & 1087 & 162 & 111 & 31.48 & 11.98 \\
$\mathsf{squar5\_261}$ & 1993 & 1298 & 1102 & 202 & 126 & 37.62 & 15.10 \\
$\mathsf{sqrt8\_260}$ & 3009 & 2104 & 1736 & 379 & 221 & 41.69 & 17.49 \\
$\mathsf{z4\_268}$ & 3073 & 2039 & 1725 & 337 & 192 & 43.03 & 15.40 \\
$\mathsf{radd\_250}$ & 3213 & 2223 & 1864 & 383 & 226 & 40.99 & 16.15 \\
$\mathsf{adr4\_197}$ & 3439 & 2359 & 1929 & 386 & 222 & 42.49 & 18.23 \\
$\mathsf{sym6\_145}$ & 3888 & 2766 & 2289 & 458 & 142 & 69.00 & 17.25 \\
$\mathsf{misex1\_241}$ & 4813 & 3202 & 2756 & 477 & 220 & 53.88 & 13.93 \\
$\mathsf{rd73\_252}$ & 5321 & 3756 & 3026 & 684 & 425 & 37.87 & 19.44 \\
$\mathsf{cycle10\_2\_110}$ & 6050 & 4333 & 3592 & 743 & 457 & 38.49 & 17.10 \\
$\mathsf{hwb6\_56}$ & 6723 & 4901 & 3976 & 894 & 435 & 51.34 & 18.87 \\
$\mathsf{square\_root\_7}$ & 7630 & 4640 & 4025 & 734 & 443 & 39.65 & 13.25 \\
$\mathsf{ham15\_107}$ & 8763 & 6119 & 5143 & 1020 & 791 & 22.45 & 15.95 \\
$\mathsf{dc2\_222}$ & 9462 & 6672 & 5600 & 1155 & 835 & 27.71 & 16.07 \\
$\mathsf{sqn\_258}$ & 10223 & 7066 & 5773 & 1240 & 775 & 37.50 & 18.30 \\
$\mathsf{inc\_237}$ & 10619 & 7232 & 6085 & 1127 & 471 & 58.21 & 15.86 \\
$\mathsf{cm85a\_209}$ & 11414 & 8144 & 6709 & 1407 & 753 & 46.48 & 17.62 \\
$\mathsf{rd84\_253}$ & 13658 & 9773 & 7798 & 2056 & 1248 & 39.30 & 20.21 \\
$\mathsf{root\_255}$ & 17159 & 11812 & 9542 & 2398 & 1823 & 23.98 & 19.22 \\
$\mathsf{co14\_215}$ & 17936 & 11652 & 9358 & 2383 & 1930 & 19.01 & 19.69 \\
$\mathsf{mlp4\_245}$ & 18852 & 13376 & 11051 & 2454 & 1518 & 38.14 & 17.38 \\
$\mathsf{urf2\_277}$ & 20112 & 15503 & 12292 & 3307 & 2590 & 21.68 & 20.71 \\
$\mathsf{sym9\_148}$ & 21504 & 14640 & 12488 & 2146 & 893 & 58.39 & 14.70 \\
$\mathsf{life\_238}$ & 22445 & 16531 & 13247 & 3251 & 1840 & 43.40 & 19.87 \\
$\mathsf{hwb7\_59}$ & 24379 & 17690 & 14203 & 3431 & 1480 & 56.86 & 19.71 \\
$\mathsf{max46\_240}$ & 27126 & 19153 & 15298 & 3807 & 2341 & 38.51 & 20.13 \\
$\mathsf{clip\_206}$ & 33827 & 24011 & 19197 & 4725 & 3091 & 34.58 & 20.05 \\
$\mathsf{sym9\_193}$ & 34881 & 25642 & 20763 & 5123 & 3589 & 29.94 & 19.03 \\
$\mathsf{9symml\_195}$ & 34881 & 25581 & 20776 & 4973 & 3460 & 30.42 & 18.78 \\
$\mathsf{dist\_223}$ & 38046 & 27164 & 21676 & 5607 & 4357 & 22.29 & 20.20 \\
$\mathsf{sao2\_257}$ & 38577 & 26191 & 21389 & 5098 & 3865 & 24.19 & 18.33 \\
$\mathsf{urf5\_280}$ & 49829 & 38073 & 31113 & 8271 & 7381 & 10.76 & 18.28 \\
$\mathsf{urf1\_278}$ & 54766 & 41840 & 34162 & 9349 & 8027 & 14.14 & 18.35 \\
$\mathsf{sym10\_262}$ & 64283 & 48065 & 39362 & 10101 & 9156 & 9.36 & 18.11 \\
$\mathsf{hwb8\_113}$ & 69380 & 51560 & 42613 & 10664 & 9217 & 13.57 & 17.35 \\
$\mathsf{urf2\_152}$ & 80480 & 58660 & 48554 & 11805 & 11275 & 4.49 & 17.23 \\
\end{longtable}
\end{center}

\scriptsize
\renewcommand{\arraystretch}{0.8}
\setlength{\tabcolsep}{13pt} 
\begin{center}
\begin{longtable}{lccccccc}
\caption{DIRSH VS SABRE 30 seconds}
\label{tab:30s} \\
\toprule
\textbf{Instance Name} & \textbf{Size} & \textbf{Sabre Depth} & \textbf{DIRSH Depth} & \textbf{Sabre Swaps} & \textbf{DIRSH Swaps} & \textbf{$\Delta$ Swap} & \textbf{$\Delta$ Depth} \\
\midrule
\endfirsthead

\toprule
\textbf{Instance Name} & \textbf{Size} & \textbf{Sabre Depth} & \textbf{DIRSH Depth} & \textbf{Sabre Swaps} & \textbf{DIRSH Swaps} & \textbf{$\Delta$ Swap} & \textbf{$\Delta$ Depth} \\
\midrule
\endhead

\bottomrule
\endfoot

\bottomrule
\endlastfoot
$\mathsf{graycode6\_47}$ & 5 & 5 & 5 & 3 & 3 & 0.00 & 0.00 \\
$\mathsf{ex1\_226}$ & 7 & 7 & 7 & 5 & 5 & 0.00 & 0.00 \\
$\mathsf{xor5\_254}$ & 7 & 7 & 7 & 5 & 5 & 0.00 & 0.00 \\
$\mathsf{4gt11\_84}$ & 18 & 15 & 13 & 4 & 4 & 0.00 & 13.33 \\
$\mathsf{ex-1\_166}$ & 19 & 16 & 15 & 3 & 3 & 0.00 & 6.25 \\
$\mathsf{ham3\_102}$ & 20 & 18 & 16 & 4 & 4 & 0.00 & 11.11 \\
$\mathsf{4mod5-v0\_20}$ & 20 & 13 & 14 & 4 & 4 & 0.00 & -7.69 \\
$\mathsf{4mod5-v1\_22}$ & 21 & 15 & 15 & 5 & 5 & 0.00 & 0.00 \\
$\mathsf{mod5d1\_63}$ & 22 & 16 & 16 & 6 & 6 & 0.00 & 0.00 \\
$\mathsf{4gt11\_83}$ & 23 & 16 & 16 & 5 & 5 & 0.00 & 0.00 \\
$\mathsf{4gt11\_82}$ & 27 & 22 & 21 & 7 & 7 & 0.00 & 4.55 \\
$\mathsf{rd32-v0\_66}$ & 34 & 29 & 25 & 8 & 8 & 0.00 & 13.79 \\
$\mathsf{4mod5-v0\_19}$ & 35 & 24 & 25 & 5 & 5 & 0.00 & -4.17 \\
$\mathsf{mod5mils\_65}$ & 35 & 24 & 24 & 6 & 6 & 0.00 & 0.00 \\
$\mathsf{4mod5-v1\_24}$ & 36 & 23 & 24 & 8 & 7 & 12.50 & -4.35 \\
$\mathsf{alu-v0\_27}$ & 36 & 29 & 25 & 9 & 8 & 11.11 & 13.79 \\
$\mathsf{rd32-v1\_68}$ & 36 & 30 & 26 & 8 & 8 & 0.00 & 13.33 \\
$\mathsf{3\_17\_13}$ & 36 & 32 & 27 & 7 & 7 & 0.00 & 15.62 \\
$\mathsf{alu-v4\_37}$ & 37 & 30 & 26 & 9 & 8 & 11.11 & 13.33 \\
$\mathsf{alu-v1\_28}$ & 37 & 26 & 26 & 7 & 7 & 0.00 & 0.00 \\
$\mathsf{alu-v3\_35}$ & 37 & 23 & 23 & 6 & 6 & 0.00 & 0.00 \\
$\mathsf{alu-v2\_33}$ & 37 & 29 & 26 & 9 & 9 & 0.00 & 10.34 \\
$\mathsf{alu-v1\_29}$ & 37 & 29 & 25 & 8 & 7 & 12.50 & 13.79 \\
$\mathsf{miller\_11}$ & 50 & 43 & 37 & 9 & 9 & 0.00 & 13.95 \\
$\mathsf{decod24-v0\_38}$ & 51 & 43 & 38 & 11 & 11 & 0.00 & 11.63 \\
$\mathsf{alu-v3\_34}$ & 52 & 33 & 36 & 7 & 7 & 0.00 & -9.09 \\
$\mathsf{decod24-v2\_43}$ & 52 & 45 & 38 & 11 & 11 & 0.00 & 15.56 \\
$\mathsf{mod5d2\_64}$ & 53 & 38 & 36 & 11 & 11 & 0.00 & 5.26 \\
$\mathsf{4gt13\_92}$ & 66 & 47 & 48 & 12 & 11 & 8.33 & -2.13 \\
$\mathsf{4gt13-v1\_93}$ & 68 & 47 & 47 & 11 & 11 & 0.00 & 0.00 \\
$\mathsf{4mod5-v0\_18}$ & 69 & 48 & 45 & 12 & 12 & 0.00 & 6.25 \\
$\mathsf{4mod5-v1\_23}$ & 69 & 48 & 47 & 12 & 12 & 0.00 & 2.08 \\
$\mathsf{one-two-three-v2\_100}$ & 69 & 49 & 43 & 13 & 9 & 30.77 & 12.24 \\
$\mathsf{one-two-three-v3\_101}$ & 70 & 47 & 46 & 11 & 11 & 0.00 & 2.13 \\
$\mathsf{4mod5-bdd\_287}$ & 70 & 46 & 43 & 8 & 8 & 0.00 & 6.52 \\
$\mathsf{qe\_qft\_4}$ & 71 & 39 & 39 & 5 & 5 & 0.00 & 0.00 \\
$\mathsf{decod24-bdd\_294}$ & 73 & 50 & 45 & 10 & 10 & 0.00 & 10.00 \\
$\mathsf{4gt5\_75}$ & 83 & 56 & 52 & 11 & 11 & 0.00 & 7.14 \\
$\mathsf{alu-bdd\_288}$ & 84 & 53 & 51 & 10 & 10 & 0.00 & 3.77 \\
$\mathsf{rd32\_270}$ & 84 & 61 & 55 & 14 & 14 & 0.00 & 9.84 \\
$\mathsf{alu-v0\_26}$ & 84 & 64 & 57 & 14 & 14 & 0.00 & 10.94 \\
$\mathsf{decod24-v1\_41}$ & 85 & 59 & 56 & 12 & 12 & 0.00 & 5.08 \\
$\mathsf{4gt5\_76}$ & 91 & 68 & 61 & 15 & 12 & 20.00 & 10.29 \\
$\mathsf{4gt13\_91}$ & 103 & 71 & 66 & 13 & 13 & 0.00 & 7.04 \\
$\mathsf{qe\_qft\_5}$ & 107 & 53 & 52 & 2 & 2 & 0.00 & 1.89 \\
$\mathsf{4gt13\_90}$ & 107 & 77 & 70 & 16 & 16 & 0.00 & 9.09 \\
$\mathsf{alu-v4\_36}$ & 115 & 80 & 75 & 14 & 14 & 0.00 & 6.25 \\
$\mathsf{4gt5\_77}$ & 131 & 86 & 79 & 17 & 14 & 17.65 & 8.14 \\
$\mathsf{rd53\_138}$ & 132 & 67 & 63 & 15 & 11 & 26.67 & 5.97 \\
$\mathsf{one-two-three-v1\_99}$ & 132 & 98 & 86 & 22 & 20 & 9.09 & 12.24 \\
$\mathsf{one-two-three-v0\_98}$ & 146 & 97 & 90 & 19 & 19 & 0.00 & 7.22 \\
$\mathsf{4gt10-v1\_81}$ & 148 & 107 & 93 & 23 & 20 & 13.04 & 13.08 \\
$\mathsf{decod24-v3\_45}$ & 150 & 104 & 92 & 23 & 17 & 26.09 & 11.54 \\
$\mathsf{aj-e11\_165}$ & 151 & 111 & 97 & 23 & 21 & 8.70 & 12.61 \\
$\mathsf{4mod7-v0\_94}$ & 162 & 106 & 103 & 19 & 15 & 21.05 & 2.83 \\
$\mathsf{alu-v2\_32}$ & 163 & 113 & 103 & 24 & 23 & 4.17 & 8.85 \\
$\mathsf{4mod7-v1\_96}$ & 164 & 111 & 104 & 21 & 19 & 9.52 & 6.31 \\
$\mathsf{mini\_alu\_305}$ & 173 & 86 & 73 & 17 & 17 & 0.00 & 15.12 \\
$\mathsf{cnt3-5\_179}$ & 175 & 75 & 66 & 26 & 22 & 15.38 & 12.00 \\
$\mathsf{mod10\_176}$ & 178 & 130 & 110 & 28 & 21 & 25.00 & 15.38 \\
$\mathsf{4gt4-v0\_80}$ & 179 & 115 & 107 & 22 & 16 & 27.27 & 6.96 \\
$\mathsf{4gt12-v0\_88}$ & 194 & 129 & 117 & 23 & 19 & 17.39 & 9.30 \\
$\mathsf{qft\_10}$ & 200 & 91 & 75 & 25 & 25 & 0.00 & 17.58 \\
$\mathsf{0410184\_169}$ & 211 & 124 & 112 & 26 & 30 & -15.38 & 9.68 \\
$\mathsf{sys6-v0\_111}$ & 215 & 94 & 85 & 21 & 20 & 4.76 & 9.57 \\
$\mathsf{4\_49\_16}$ & 217 & 159 & 134 & 34 & 26 & 23.53 & 15.72 \\
$\mathsf{4gt12-v1\_89}$ & 228 & 151 & 136 & 24 & 13 & 45.83 & 9.93 \\
$\mathsf{rd73\_140}$ & 230 & 121 & 103 & 27 & 24 & 11.11 & 14.88 \\
$\mathsf{4gt4-v0\_79}$ & 231 & 150 & 141 & 23 & 18 & 21.74 & 6.00 \\
$\mathsf{hwb4\_49}$ & 233 & 178 & 150 & 38 & 32 & 15.79 & 15.73 \\
$\mathsf{4gt4-v0\_78}$ & 235 & 154 & 146 & 22 & 20 & 9.09 & 5.19 \\
$\mathsf{mod10\_171}$ & 244 & 177 & 149 & 37 & 32 & 13.51 & 15.82 \\
$\mathsf{4gt12-v0\_87}$ & 247 & 147 & 142 & 18 & 17 & 5.56 & 3.40 \\
$\mathsf{4gt12-v0\_86}$ & 251 & 152 & 147 & 17 & 19 & -11.76 & 3.29 \\
$\mathsf{4gt4-v0\_72}$ & 258 & 156 & 141 & 22 & 16 & 27.27 & 9.62 \\
$\mathsf{sym6\_316}$ & 270 & 162 & 143 & 41 & 33 & 19.51 & 11.73 \\
$\mathsf{4gt4-v1\_74}$ & 273 & 186 & 165 & 35 & 22 & 37.14 & 11.29 \\
$\mathsf{rd53\_311}$ & 275 & 154 & 137 & 38 & 34 & 10.53 & 11.04 \\
$\mathsf{mini-alu\_167}$ & 288 & 209 & 175 & 49 & 42 & 14.29 & 16.27 \\
$\mathsf{one-two-three-v0\_97}$ & 290 & 211 & 177 & 46 & 38 & 17.39 & 16.11 \\
$\mathsf{rd53\_135}$ & 296 & 202 & 168 & 42 & 29 & 30.95 & 16.83 \\
$\mathsf{ham7\_104}$ & 320 & 215 & 197 & 35 & 24 & 31.43 & 8.37 \\
$\mathsf{sym9\_146}$ & 328 & 182 & 146 & 50 & 42 & 16.00 & 19.78 \\
$\mathsf{decod24-enable\_126}$ & 338 & 221 & 199 & 31 & 32 & -3.23 & 9.95 \\
$\mathsf{mod8-10\_178}$ & 342 & 221 & 204 & 26 & 20 & 23.08 & 7.69 \\
$\mathsf{rd84\_142}$ & 343 & 161 & 128 & 56 & 50 & 10.71 & 20.50 \\
$\mathsf{4gt4-v0\_73}$ & 395 & 276 & 241 & 43 & 34 & 20.93 & 12.68 \\
$\mathsf{ex3\_229}$ & 403 & 272 & 235 & 38 & 25 & 34.21 & 13.60 \\
$\mathsf{mod8-10\_177}$ & 440 & 312 & 266 & 53 & 31 & 41.51 & 14.74 \\
$\mathsf{alu-v2\_31}$ & 451 & 337 & 283 & 75 & 69 & 8.00 & 16.02 \\
$\mathsf{C17\_204}$ & 467 & 295 & 264 & 37 & 34 & 8.11 & 10.51 \\
$\mathsf{rd53\_131}$ & 469 & 320 & 273 & 57 & 34 & 40.35 & 14.69 \\
$\mathsf{ising\_model\_10}$ & 480 & 92 & 71 & 6 & 6 & 0.00 & 22.83 \\
$\mathsf{cnt3-5\_180}$ & 485 & 290 & 228 & 70 & 58 & 17.14 & 21.38 \\
$\mathsf{alu-v2\_30}$ & 504 & 358 & 306 & 62 & 42 & 32.26 & 14.53 \\
$\mathsf{qft\_16}$ & 512 & 216 & 137 & 64 & 70 & -9.38 & 36.57 \\
$\mathsf{mod5adder\_127}$ & 555 & 384 & 319 & 63 & 44 & 30.16 & 16.93 \\
$\mathsf{rd53\_133}$ & 580 & 395 & 350 & 56 & 71 & -26.79 & 11.39 \\
$\mathsf{majority\_239}$ & 612 & 425 & 360 & 66 & 58 & 12.12 & 15.29 \\
$\mathsf{ex2\_227}$ & 631 & 446 & 373 & 77 & 69 & 10.39 & 16.37 \\
$\mathsf{ising\_model\_13}$ & 633 & 92 & 86 & 8 & 16 & -100.00 & 6.52 \\
$\mathsf{cm82a\_208}$ & 650 & 416 & 356 & 69 & 34 & 50.72 & 14.42 \\
$\mathsf{sf\_276}$ & 778 & 490 & 460 & 44 & 37 & 15.91 & 6.12 \\
$\mathsf{sf\_274}$ & 781 & 498 & 452 & 51 & 32 & 37.25 & 9.24 \\
$\mathsf{ising\_model\_16}$ & 786 & 105 & 88 & 14 & 20 & -42.86 & 16.19 \\
$\mathsf{con1\_216}$ & 954 & 602 & 538 & 88 & 67 & 23.86 & 10.63 \\
$\mathsf{wim\_266}$ & 986 & 646 & 540 & 118 & 58 & 50.85 & 16.41 \\
$\mathsf{rd53\_130}$ & 1043 & 751 & 605 & 140 & 78 & 44.29 & 19.44 \\
$\mathsf{f2\_232}$ & 1206 & 860 & 711 & 167 & 88 & 47.31 & 17.33 \\
$\mathsf{cm152a\_212}$ & 1221 & 827 & 721 & 141 & 92 & 34.75 & 12.82 \\
$\mathsf{rd53\_251}$ & 1291 & 911 & 763 & 163 & 95 & 41.72 & 16.25 \\
$\mathsf{hwb5\_53}$ & 1336 & 974 & 798 & 176 & 111 & 36.93 & 18.07 \\
$\mathsf{pm1\_249}$ & 1776 & 1166 & 985 & 189 & 113 & 40.21 & 15.52 \\
$\mathsf{cm42a\_207}$ & 1776 & 1161 & 987 & 189 & 120 & 36.51 & 14.99 \\
$\mathsf{dc1\_220}$ & 1914 & 1235 & 1084 & 162 & 109 & 32.72 & 12.23 \\
$\mathsf{squar5\_261}$ & 1993 & 1286 & 1101 & 188 & 117 & 37.77 & 14.39 \\
$\mathsf{sqrt8\_260}$ & 3009 & 2104 & 1744 & 379 & 220 & 41.95 & 17.11 \\
$\mathsf{z4\_268}$ & 3073 & 2037 & 1720 & 321 & 202 & 37.07 & 15.56 \\
$\mathsf{radd\_250}$ & 3213 & 2223 & 1863 & 383 & 220 & 42.56 & 16.19 \\
$\mathsf{adr4\_197}$ & 3439 & 2332 & 1939 & 386 & 249 & 35.49 & 16.85 \\
$\mathsf{sym6\_145}$ & 3888 & 2734 & 2299 & 429 & 140 & 67.37 & 15.91 \\
$\mathsf{misex1\_241}$ & 4813 & 3202 & 2751 & 475 & 190 & 60.00 & 14.08 \\
$\mathsf{rd73\_252}$ & 5321 & 3756 & 3023 & 684 & 413 & 39.62 & 19.52 \\
$\mathsf{cycle10\_2\_110}$ & 6050 & 4312 & 3585 & 743 & 469 & 36.88 & 16.86 \\
$\mathsf{hwb6\_56}$ & 6723 & 4901 & 3973 & 894 & 436 & 51.23 & 18.93 \\
$\mathsf{square\_root\_7}$ & 7630 & 4640 & 4020 & 734 & 419 & 42.92 & 13.36 \\
$\mathsf{ham15\_107}$ & 8763 & 6119 & 5127 & 1020 & 717 & 29.71 & 16.21 \\
$\mathsf{dc2\_222}$ & 9462 & 6605 & 5558 & 1150 & 807 & 29.83 & 15.85 \\
$\mathsf{sqn\_258}$ & 10223 & 7066 & 5753 & 1240 & 723 & 41.69 & 18.58 \\
$\mathsf{inc\_237}$ & 10619 & 7209 & 6074 & 1040 & 497 & 52.21 & 15.74 \\
$\mathsf{cm85a\_209}$ & 11414 & 8048 & 6664 & 1358 & 745 & 45.14 & 17.20 \\
$\mathsf{rd84\_253}$ & 13658 & 9773 & 7746 & 2001 & 1267 & 36.68 & 20.74 \\
$\mathsf{root\_255}$ & 17159 & 11812 & 9464 & 2398 & 1647 & 31.32 & 19.88 \\
$\mathsf{co14\_215}$ & 17936 & 11652 & 9319 & 2383 & 1773 & 25.60 & 20.02 \\
$\mathsf{mlp4\_245}$ & 18852 & 13322 & 10997 & 2449 & 1617 & 33.97 & 17.45 \\
$\mathsf{urf2\_277}$ & 20112 & 15503 & 12251 & 3307 & 2478 & 25.07 & 20.98 \\
$\mathsf{sym9\_148}$ & 21504 & 14640 & 12444 & 2146 & 816 & 61.98 & 15.00 \\
$\mathsf{life\_238}$ & 22445 & 16528 & 13233 & 3214 & 1894 & 41.07 & 19.94 \\
$\mathsf{hwb7\_59}$ & 24379 & 17653 & 14154 & 3431 & 1368 & 60.13 & 19.82 \\
$\mathsf{max46\_240}$ & 27126 & 19153 & 15236 & 3807 & 2274 & 40.27 & 20.45 \\
$\mathsf{clip\_206}$ & 33827 & 23920 & 19075 & 4725 & 3096 & 34.48 & 20.26 \\
$\mathsf{sym9\_193}$ & 34881 & 25642 & 20750 & 5123 & 3354 & 34.53 & 19.08 \\
$\mathsf{9symml\_195}$ & 34881 & 25571 & 20679 & 4973 & 3461 & 30.40 & 19.13 \\
$\mathsf{dist\_223}$ & 38046 & 27013 & 21597 & 5555 & 4247 & 23.55 & 20.05 \\
$\mathsf{sao2\_257}$ & 38577 & 26108 & 21237 & 5098 & 3783 & 25.79 & 18.66 \\
$\mathsf{urf5\_280}$ & 49829 & 37919 & 30965 & 8183 & 7000 & 14.46 & 18.34 \\
$\mathsf{urf1\_278}$ & 54766 & 41564 & 34157 & 9077 & 7880 & 13.19 & 17.82 \\
$\mathsf{sym10\_262}$ & 64283 & 47886 & 39270 & 9786 & 7924 & 19.03 & 17.99 \\
$\mathsf{hwb8\_113}$ & 69380 & 51560 & 42376 & 10511 & 8910 & 15.23 & 17.81 \\
$\mathsf{urf2\_152}$ & 80480 & 58625 & 48449 & 11776 & 10797 & 8.31 & 17.36 \\
\end{longtable}
\end{center}

\scriptsize
\renewcommand{\arraystretch}{0.8}
\setlength{\tabcolsep}{13pt} 
\begin{center}
\begin{longtable}{lccccccc}
\caption{DIRSH VS SABRE 60 seconds}
\label{tab:60s} \\
\toprule
\textbf{Instance Name} & \textbf{Size} & \textbf{Sabre Depth} & \textbf{DIRSH Depth} & \textbf{Sabre Swaps} & \textbf{DIRSH Swaps} & \textbf{$\Delta$ Swap} & \textbf{$\Delta$ Depth} \\
\midrule
\endfirsthead

\toprule
\textbf{Instance Name} & \textbf{Size} & \textbf{Sabre Depth} & \textbf{DIRSH Depth} & \textbf{Sabre Swaps} & \textbf{DIRSH Swaps} & \textbf{$\Delta$ Swap} & \textbf{$\Delta$ Depth} \\
\midrule
\endhead

\bottomrule
\endfoot

\bottomrule
\endlastfoot
$\mathsf{graycode6\_47}$ & 5 & 5 & 5 & 3 & 3 & 0.00 & 0.00 \\
$\mathsf{ex1\_226}$ & 7 & 7 & 7 & 5 & 5 & 0.00 & 0.00 \\
$\mathsf{xor5\_254}$ & 7 & 7 & 7 & 5 & 5 & 0.00 & 0.00 \\
$\mathsf{4gt11\_84}$ & 18 & 15 & 13 & 4 & 4 & 0.00 & 13.33 \\
$\mathsf{ex-1\_166}$ & 19 & 16 & 15 & 3 & 3 & 0.00 & 6.25 \\
$\mathsf{ham3\_102}$ & 20 & 18 & 16 & 4 & 4 & 0.00 & 11.11 \\
$\mathsf{4mod5-v0\_20}$ & 20 & 13 & 14 & 4 & 4 & 0.00 & -7.69 \\
$\mathsf{4mod5-v1\_22}$ & 21 & 15 & 15 & 5 & 5 & 0.00 & 0.00 \\
$\mathsf{mod5d1\_63}$ & 22 & 16 & 16 & 6 & 6 & 0.00 & 0.00 \\
$\mathsf{4gt11\_83}$ & 23 & 16 & 16 & 5 & 5 & 0.00 & 0.00 \\
$\mathsf{4gt11\_82}$ & 27 & 22 & 21 & 7 & 7 & 0.00 & 4.55 \\
$\mathsf{rd32-v0\_66}$ & 34 & 29 & 25 & 8 & 8 & 0.00 & 13.79 \\
$\mathsf{4mod5-v0\_19}$ & 35 & 24 & 25 & 5 & 5 & 0.00 & -4.17 \\
$\mathsf{mod5mils\_65}$ & 35 & 24 & 24 & 6 & 6 & 0.00 & 0.00 \\
$\mathsf{4mod5-v1\_24}$ & 36 & 23 & 24 & 8 & 7 & 12.50 & -4.35 \\
$\mathsf{alu-v0\_27}$ & 36 & 29 & 25 & 9 & 8 & 11.11 & 13.79 \\
$\mathsf{rd32-v1\_68}$ & 36 & 30 & 26 & 8 & 8 & 0.00 & 13.33 \\
$\mathsf{3\_17\_13}$ & 36 & 32 & 27 & 7 & 7 & 0.00 & 15.62 \\
$\mathsf{alu-v4\_37}$ & 37 & 30 & 26 & 9 & 8 & 11.11 & 13.33 \\
$\mathsf{alu-v1\_28}$ & 37 & 26 & 26 & 7 & 7 & 0.00 & 0.00 \\
$\mathsf{alu-v3\_35}$ & 37 & 23 & 23 & 6 & 6 & 0.00 & 0.00 \\
$\mathsf{alu-v2\_33}$ & 37 & 29 & 26 & 9 & 9 & 0.00 & 10.34 \\
$\mathsf{alu-v1\_29}$ & 37 & 29 & 25 & 8 & 7 & 12.50 & 13.79 \\
$\mathsf{miller\_11}$ & 50 & 43 & 37 & 9 & 9 & 0.00 & 13.95 \\
$\mathsf{decod24-v0\_38}$ & 51 & 43 & 38 & 11 & 11 & 0.00 & 11.63 \\
$\mathsf{alu-v3\_34}$ & 52 & 33 & 36 & 7 & 7 & 0.00 & -9.09 \\
$\mathsf{decod24-v2\_43}$ & 52 & 45 & 38 & 11 & 11 & 0.00 & 15.56 \\
$\mathsf{mod5d2\_64}$ & 53 & 38 & 36 & 11 & 11 & 0.00 & 5.26 \\
$\mathsf{4gt13\_92}$ & 66 & 47 & 48 & 12 & 11 & 8.33 & -2.13 \\
$\mathsf{4gt13-v1\_93}$ & 68 & 47 & 47 & 11 & 11 & 0.00 & 0.00 \\
$\mathsf{4mod5-v0\_18}$ & 69 & 48 & 45 & 12 & 12 & 0.00 & 6.25 \\
$\mathsf{4mod5-v1\_23}$ & 69 & 48 & 47 & 12 & 12 & 0.00 & 2.08 \\
$\mathsf{one-two-three-v2\_100}$ & 69 & 49 & 43 & 13 & 9 & 30.77 & 12.24 \\
$\mathsf{one-two-three-v3\_101}$ & 70 & 47 & 46 & 11 & 11 & 0.00 & 2.13 \\
$\mathsf{4mod5-bdd\_287}$ & 70 & 46 & 43 & 8 & 8 & 0.00 & 6.52 \\
$\mathsf{qe\_qft\_4}$ & 71 & 39 & 39 & 5 & 5 & 0.00 & 0.00 \\
$\mathsf{decod24-bdd\_294}$ & 73 & 50 & 45 & 10 & 10 & 0.00 & 10.00 \\
$\mathsf{4gt5\_75}$ & 83 & 56 & 52 & 11 & 11 & 0.00 & 7.14 \\
$\mathsf{alu-bdd\_288}$ & 84 & 53 & 51 & 10 & 10 & 0.00 & 3.77 \\
$\mathsf{rd32\_270}$ & 84 & 60 & 55 & 14 & 14 & 0.00 & 8.33 \\
$\mathsf{alu-v0\_26}$ & 84 & 64 & 57 & 14 & 14 & 0.00 & 10.94 \\
$\mathsf{decod24-v1\_41}$ & 85 & 59 & 56 & 12 & 12 & 0.00 & 5.08 \\
$\mathsf{4gt5\_76}$ & 91 & 68 & 61 & 15 & 12 & 20.00 & 10.29 \\
$\mathsf{4gt13\_91}$ & 103 & 69 & 66 & 13 & 13 & 0.00 & 4.35 \\
$\mathsf{qe\_qft\_5}$ & 107 & 53 & 52 & 2 & 2 & 0.00 & 1.89 \\
$\mathsf{4gt13\_90}$ & 107 & 77 & 70 & 16 & 16 & 0.00 & 9.09 \\
$\mathsf{alu-v4\_36}$ & 115 & 80 & 75 & 14 & 14 & 0.00 & 6.25 \\
$\mathsf{4gt5\_77}$ & 131 & 85 & 79 & 16 & 14 & 12.50 & 7.06 \\
$\mathsf{rd53\_138}$ & 132 & 66 & 63 & 15 & 11 & 26.67 & 4.55 \\
$\mathsf{one-two-three-v1\_99}$ & 132 & 97 & 86 & 22 & 20 & 9.09 & 11.34 \\
$\mathsf{one-two-three-v0\_98}$ & 146 & 97 & 90 & 19 & 19 & 0.00 & 7.22 \\
$\mathsf{4gt10-v1\_81}$ & 148 & 104 & 93 & 23 & 20 & 13.04 & 10.58 \\
$\mathsf{decod24-v3\_45}$ & 150 & 103 & 92 & 20 & 17 & 15.00 & 10.68 \\
$\mathsf{aj-e11\_165}$ & 151 & 109 & 97 & 23 & 21 & 8.70 & 11.01 \\
$\mathsf{4mod7-v0\_94}$ & 162 & 106 & 103 & 19 & 15 & 21.05 & 2.83 \\
$\mathsf{alu-v2\_32}$ & 163 & 113 & 103 & 23 & 23 & 0.00 & 8.85 \\
$\mathsf{4mod7-v1\_96}$ & 164 & 111 & 104 & 21 & 19 & 9.52 & 6.31 \\
$\mathsf{mini\_alu\_305}$ & 173 & 86 & 73 & 17 & 16 & 5.88 & 15.12 \\
$\mathsf{cnt3-5\_179}$ & 175 & 75 & 66 & 26 & 22 & 15.38 & 12.00 \\
$\mathsf{mod10\_176}$ & 178 & 130 & 110 & 27 & 21 & 22.22 & 15.38 \\
$\mathsf{4gt4-v0\_80}$ & 179 & 113 & 107 & 20 & 16 & 20.00 & 5.31 \\
$\mathsf{4gt12-v0\_88}$ & 194 & 129 & 117 & 23 & 19 & 17.39 & 9.30 \\
$\mathsf{qft\_10}$ & 200 & 90 & 75 & 25 & 25 & 0.00 & 16.67 \\
$\mathsf{0410184\_169}$ & 211 & 124 & 112 & 26 & 30 & -15.38 & 9.68 \\
$\mathsf{sys6-v0\_111}$ & 215 & 94 & 85 & 21 & 19 & 9.52 & 9.57 \\
$\mathsf{4\_49\_16}$ & 217 & 156 & 134 & 34 & 26 & 23.53 & 14.10 \\
$\mathsf{4gt12-v1\_89}$ & 228 & 151 & 136 & 24 & 13 & 45.83 & 9.93 \\
$\mathsf{rd73\_140}$ & 230 & 121 & 103 & 27 & 24 & 11.11 & 14.88 \\
$\mathsf{4gt4-v0\_79}$ & 231 & 149 & 141 & 23 & 18 & 21.74 & 5.37 \\
$\mathsf{hwb4\_49}$ & 233 & 178 & 148 & 38 & 32 & 15.79 & 16.85 \\
$\mathsf{4gt4-v0\_78}$ & 235 & 151 & 146 & 22 & 19 & 13.64 & 3.31 \\
$\mathsf{mod10\_171}$ & 244 & 171 & 149 & 34 & 32 & 5.88 & 12.87 \\
$\mathsf{4gt12-v0\_87}$ & 247 & 147 & 142 & 18 & 17 & 5.56 & 3.40 \\
$\mathsf{4gt12-v0\_86}$ & 251 & 152 & 147 & 17 & 19 & -11.76 & 3.29 \\
$\mathsf{4gt4-v0\_72}$ & 258 & 152 & 141 & 20 & 16 & 20.00 & 7.24 \\
$\mathsf{sym6\_316}$ & 270 & 162 & 143 & 39 & 32 & 17.95 & 11.73 \\
$\mathsf{4gt4-v1\_74}$ & 273 & 186 & 165 & 35 & 22 & 37.14 & 11.29 \\
$\mathsf{rd53\_311}$ & 275 & 151 & 137 & 38 & 34 & 10.53 & 9.27 \\
$\mathsf{mini-alu\_167}$ & 288 & 209 & 175 & 49 & 39 & 20.41 & 16.27 \\
$\mathsf{one-two-three-v0\_97}$ & 290 & 211 & 176 & 42 & 38 & 9.52 & 16.59 \\
$\mathsf{rd53\_135}$ & 296 & 199 & 168 & 39 & 29 & 25.64 & 15.58 \\
$\mathsf{ham7\_104}$ & 320 & 210 & 197 & 32 & 22 & 31.25 & 6.19 \\
$\mathsf{sym9\_146}$ & 328 & 182 & 145 & 40 & 42 & -5.00 & 20.33 \\
$\mathsf{decod24-enable\_126}$ & 338 & 221 & 199 & 31 & 32 & -3.23 & 9.95 \\
$\mathsf{mod8-10\_178}$ & 342 & 221 & 203 & 26 & 20 & 23.08 & 8.14 \\
$\mathsf{rd84\_142}$ & 343 & 160 & 127 & 56 & 48 & 14.29 & 20.62 \\
$\mathsf{4gt4-v0\_73}$ & 395 & 274 & 241 & 43 & 29 & 32.56 & 12.04 \\
$\mathsf{ex3\_229}$ & 403 & 267 & 235 & 34 & 25 & 26.47 & 11.99 \\
$\mathsf{mod8-10\_177}$ & 440 & 304 & 262 & 44 & 30 & 31.82 & 13.82 \\
$\mathsf{alu-v2\_31}$ & 451 & 320 & 283 & 70 & 66 & 5.71 & 11.56 \\
$\mathsf{C17\_204}$ & 467 & 289 & 264 & 33 & 31 & 6.06 & 8.65 \\
$\mathsf{rd53\_131}$ & 469 & 312 & 273 & 50 & 34 & 32.00 & 12.50 \\
$\mathsf{ising\_model\_10}$ & 480 & 92 & 71 & 6 & 6 & 0.00 & 22.83 \\
$\mathsf{cnt3-5\_180}$ & 485 & 288 & 228 & 69 & 57 & 17.39 & 20.83 \\
$\mathsf{alu-v2\_30}$ & 504 & 355 & 306 & 62 & 55 & 11.29 & 13.80 \\
$\mathsf{qft\_16}$ & 512 & 216 & 135 & 64 & 74 & -15.62 & 37.50 \\
$\mathsf{mod5adder\_127}$ & 555 & 372 & 319 & 51 & 44 & 13.73 & 14.25 \\
$\mathsf{rd53\_133}$ & 580 & 395 & 353 & 56 & 71 & -26.79 & 10.63 \\
$\mathsf{majority\_239}$ & 612 & 425 & 359 & 66 & 48 & 27.27 & 15.53 \\
$\mathsf{ex2\_227}$ & 631 & 441 & 373 & 77 & 59 & 23.38 & 15.42 \\
$\mathsf{ising\_model\_13}$ & 633 & 92 & 89 & 8 & 9 & -12.50 & 3.26 \\
$\mathsf{cm82a\_208}$ & 650 & 408 & 356 & 69 & 34 & 50.72 & 12.75 \\
$\mathsf{sf\_276}$ & 778 & 490 & 460 & 44 & 37 & 15.91 & 6.12 \\
$\mathsf{sf\_274}$ & 781 & 467 & 452 & 27 & 30 & -11.11 & 3.21 \\
$\mathsf{ising\_model\_16}$ & 786 & 104 & 92 & 13 & 15 & -15.38 & 11.54 \\
$\mathsf{con1\_216}$ & 954 & 602 & 535 & 88 & 67 & 23.86 & 11.13 \\
$\mathsf{wim\_266}$ & 986 & 646 & 540 & 118 & 61 & 48.31 & 16.41 \\
$\mathsf{rd53\_130}$ & 1043 & 747 & 605 & 140 & 81 & 42.14 & 19.01 \\
$\mathsf{f2\_232}$ & 1206 & 850 & 708 & 158 & 88 & 44.30 & 16.71 \\
$\mathsf{cm152a\_212}$ & 1221 & 827 & 722 & 141 & 90 & 36.17 & 12.70 \\
$\mathsf{rd53\_251}$ & 1291 & 910 & 756 & 163 & 88 & 46.01 & 16.92 \\
$\mathsf{hwb5\_53}$ & 1336 & 974 & 800 & 176 & 95 & 46.02 & 17.86 \\
$\mathsf{pm1\_249}$ & 1776 & 1153 & 981 & 182 & 113 & 37.91 & 14.92 \\
$\mathsf{cm42a\_207}$ & 1776 & 1161 & 995 & 187 & 111 & 40.64 & 14.30 \\
$\mathsf{dc1\_220}$ & 1914 & 1235 & 1088 & 162 & 110 & 32.10 & 11.90 \\
$\mathsf{squar5\_261}$ & 1993 & 1255 & 1103 & 172 & 120 & 30.23 & 12.11 \\
$\mathsf{sqrt8\_260}$ & 3009 & 2103 & 1743 & 371 & 200 & 46.09 & 17.12 \\
$\mathsf{z4\_268}$ & 3073 & 2019 & 1725 & 311 & 189 & 39.23 & 14.56 \\
$\mathsf{radd\_250}$ & 3213 & 2223 & 1855 & 383 & 207 & 45.95 & 16.55 \\
$\mathsf{adr4\_197}$ & 3439 & 2276 & 1921 & 386 & 219 & 43.26 & 15.60 \\
$\mathsf{sym6\_145}$ & 3888 & 2734 & 2303 & 429 & 141 & 67.13 & 15.76 \\
$\mathsf{misex1\_241}$ & 4813 & 3202 & 2752 & 460 & 153 & 66.74 & 14.05 \\
$\mathsf{rd73\_252}$ & 5321 & 3756 & 3010 & 684 & 400 & 41.52 & 19.86 \\
$\mathsf{cycle10\_2\_110}$ & 6050 & 4312 & 3550 & 743 & 438 & 41.05 & 17.67 \\
$\mathsf{hwb6\_56}$ & 6723 & 4883 & 3929 & 894 & 426 & 52.35 & 19.54 \\
$\mathsf{square\_root\_7}$ & 7630 & 4631 & 4010 & 732 & 350 & 52.19 & 13.41 \\
$\mathsf{ham15\_107}$ & 8763 & 6099 & 5094 & 1017 & 695 & 31.66 & 16.48 \\
$\mathsf{dc2\_222}$ & 9462 & 6581 & 5573 & 1117 & 731 & 34.56 & 15.32 \\
$\mathsf{sqn\_258}$ & 10223 & 7066 & 5734 & 1240 & 659 & 46.85 & 18.85 \\
$\mathsf{inc\_237}$ & 10619 & 7085 & 6060 & 1040 & 412 & 60.38 & 14.47 \\
$\mathsf{cm85a\_209}$ & 11414 & 8048 & 6662 & 1358 & 696 & 48.75 & 17.22 \\
$\mathsf{rd84\_253}$ & 13658 & 9773 & 7713 & 2001 & 1111 & 44.48 & 21.08 \\
$\mathsf{root\_255}$ & 17159 & 11659 & 9426 & 2267 & 1532 & 32.42 & 19.15 \\
$\mathsf{co14\_215}$ & 17936 & 11652 & 9305 & 2383 & 1879 & 21.15 & 20.14 \\
$\mathsf{mlp4\_245}$ & 18852 & 13322 & 10931 & 2449 & 1457 & 40.51 & 17.95 \\
$\mathsf{urf2\_277}$ & 20112 & 15439 & 12219 & 3285 & 2489 & 24.23 & 20.86 \\
$\mathsf{sym9\_148}$ & 21504 & 14591 & 12442 & 2080 & 870 & 58.17 & 14.73 \\
$\mathsf{life\_238}$ & 22445 & 16528 & 13145 & 3214 & 1667 & 48.13 & 20.47 \\
$\mathsf{hwb7\_59}$ & 24379 & 17459 & 14124 & 3300 & 1357 & 58.88 & 19.10 \\
$\mathsf{max46\_240}$ & 27126 & 19083 & 15179 & 3807 & 2103 & 44.76 & 20.46 \\
$\mathsf{clip\_206}$ & 33827 & 23864 & 19050 & 4690 & 2700 & 42.43 & 20.17 \\
$\mathsf{sym9\_193}$ & 34881 & 25528 & 20610 & 5069 & 3114 & 38.57 & 19.27 \\
$\mathsf{9symml\_195}$ & 34881 & 25571 & 20587 & 4973 & 3035 & 38.97 & 19.49 \\
$\mathsf{dist\_223}$ & 38046 & 27013 & 21406 & 5555 & 3857 & 30.57 & 20.76 \\
$\mathsf{sao2\_257}$ & 38577 & 25610 & 21066 & 4645 & 3386 & 27.10 & 17.74 \\
$\mathsf{urf5\_280}$ & 49829 & 37919 & 30817 & 8183 & 6815 & 16.72 & 18.73 \\
$\mathsf{urf1\_278}$ & 54766 & 41564 & 33873 & 9077 & 7450 & 17.92 & 18.50 \\
$\mathsf{sym10\_262}$ & 64283 & 47812 & 38948 & 9738 & 7821 & 19.69 & 18.54 \\
$\mathsf{hwb8\_113}$ & 69380 & 51560 & 42182 & 10511 & 8238 & 21.62 & 18.19 \\
$\mathsf{urf2\_152}$ & 80480 & 58583 & 48205 & 11776 & 10446 & 11.29 & 17.72 \\
\end{longtable}
\end{center}

\section*{Acknowledgments}
M. Baioletti acknowledges the financial support from the European Union - NextGenerationEU, Mission 4, Component 2, under the Italian Ministry of University and Research (MUR) National Innovation Ecosystem grant ECS00000041 - VITALITY - CUP J97G22000170005.

\bibliographystyle{ieeetr}
\bibliography{qai}

\end{document}